\documentclass[acmsmall]{acmart}

\usepackage{booktabs} % For formal tables
\usepackage{hyperref}   %gws 20170930
\usepackage{url}  %gws 20170930

\usepackage{multirow}  %gws 20170930
\usepackage{color}  %gws 20170930
\usepackage{diagbox}
\usepackage{amsmath}
\usepackage{arydshln} % gws, 20180202
\usepackage{color} % gws, 20180120
\usepackage{epstopdf}

\newtheorem{property}{Property}

\usepackage[linesnumbered,ruled]{algorithm2e} % For algorithms

\SetAlFnt{\small}
\SetAlCapFnt{\small}
\SetAlCapNameFnt{\small}
\SetAlCapHSkip{0pt}
\IncMargin{-\parindent}

\AtBeginDocument{%
  }

\setcopyright{acmcopyright}
\copyrightyear{2022}
\acmYear{2022}
\acmDOI{0000001.0000001}

\acmJournal{JACM}

\begin{document}

% \linenumbers
\title{A Generic Algorithm for Top-$K$ On-Shelf Utility Mining}

\author{Jiahui Chen}
\email{csjhchen@gmail.com}
\affiliation{%
  \institution{Guangdong University of Technology}
  \city{Guangzhou}
  \state{Guangdong}
  \country{China}
  %\postcode{}
}

\author{Xu Guo}
\affiliation{%
  \institution{Guangdong University of Technology}
  \city{Guangzhou}
  \state{Guangdong}
  \country{China}
}
\email{csxuguo@gmail.com}

\author{Wensheng Gan}
\authornote{This is the corresponding author}
\email{wsgan001@gmail.com}
\affiliation{%
  \institution{Jinan University}
  \city{Guangzhou}
  \state{Guangdong}
  \country{China}
}
%\affiliation{%
%  \institution{Pazhou Lab}
%  \city{Guangzhou}
%  \state{Guangdong}
%  \country{China}
%  \postcode{510330}
%}

\author{Shichen Wan}
\affiliation{%
  \institution{Guangdong University of Technology}
  \city{Guangzhou}
  \state{Guangdong}
  \country{China}
  %\postcode{}
}
\email{scwan1998@gmail.com}

\author{Philip S. Yu}
\affiliation{%
	\institution{University of Illinois at Chicago}
	\city{Chicago}
	\state{IL}
	\country{USA}
	\postcode{60607}
}
\email{psyu@uic.edu}

\renewcommand{\shortauthors}{Jiahui Chen et al.}

\begin{abstract}
  On-shelf utility mining (OSUM) is an emerging research direction in data mining. It aims to discover itemsets that have high relative utility in their selling time period. Compared with traditional utility mining, OSUM can find more practical and meaningful patterns in real-life applications. However, there is a major drawback to traditional OSUM. For normal users, it is hard to define a minimum threshold \textit{minutil} for mining the right amount of on-shelf high utility itemsets. On one hand, if the threshold is set too high, the number of patterns would not be enough. On the other hand, if the threshold is set too low, too many patterns will be discovered and cause an unnecessary waste of time and memory consumption. To address this issue, the user usually directly specifies a parameter $k$, where only the top-$k$ high relative utility itemsets would be considered. Therefore, in this paper, we propose a generic algorithm named TOIT for mining Top-$k$ On-shelf hIgh-utility paTterns to solve this problem. TOIT applies a novel strategy to raise the \textit{minutil} based on the on-shelf datasets. Besides, two novel upper-bound strategies named subtree utility and local utility are applied to prune the search space. By adopting the strategies mentioned above, the TOIT algorithm can narrow the search space as early as possible, improve the mining efficiency, and reduce the memory consumption, so it can obtain better performance than other algorithms. A series of experiments have been conducted on real datasets with different styles to compare the effects with the state-of-the-art KOSHU algorithm. The experimental results showed that TOIT outperforms KOSHU in both running time and memory consumption.
\end{abstract}





\maketitle

\section{Introduction}

In recent years, accompanied by the various increasing investigations of frequency-based pattern mining (FPM) \cite{agarwal1994fast, han2000mining,fournier2017survey,wu2021ntp} in the field of data mining, utility-driven pattern mining, also called high utility pattern mining (HUPM) \cite{ahmed2009efficient, fournier2014fhm, gan2019huopm, gan2019survey} is in the spotlight. The aim of HUPM is to discover a set of patterns (e.g., itemsets) which generates high utility no less than the utility threshold set by the user. In contrast to the FPM, the HUPM allows each item to appear multiple times in a record (e.g., a transaction) and has its own value. HUPM is more suitable for the current situation, and can discover more practical and useful knowledge from databases that have rich information. Let us consider a special situation where the supermarket sells some discount products (which can be unprofitable) to attract consumers to the supermarket. This situation not only requires HUPM to handle the positive utility items, but also the negative utility items at the same time. In FPM, there exists a famous downward closure property, where the support of any superset of an itemset will monotonically decrease or remain unchanged. However, this property can not be simply applied to HUPM, because the utility of an itemset in HUPM is neither monotonic nor anti-monotonic \cite{gan2019survey}. More precisely, although an itemset has high-utility, there may still exist some superset or subset that does not exceed the \textit{minutil}.

Experts have carried out a lot of studies on HUPM and have proposed many useful and efficient algorithms. Most of the existing HUPM algorithms can be mainly divided into two categories. One is called the two-phase-based algorithms, such as PB \cite{lan2014eff}, Two-Phase \cite{liu2005two}, BAHUI \cite{song2014bahui}, UP-Growth \cite{tseng2010up}, and UP-Growth+ \cite{tseng2012efficient}. And the other one is called the single-phase algorithms, such as HUI-Miner \cite{liu2012direct} and {$\rm d^2HUP$} \cite{liu2012direct}. Generally, for all items in HUPM, the existing algorithms usually assume that they have the same on-shelf time periods, which is irrational for real applications. Based on this premise, there will be a bias toward the product that will not be sold all the time. For example, the \{\textit{ice cream}, \textit{cold drink}\} combination is selling well in summer, but not in winter. If all the products were considered to have the same on-shelf time period, the combination of some products that were only sold well in a certain sub-period of time would be considered a low-utility itemset. Thus, the time-related information is necessary and meaningful to be considered. Consequently, the problem of on-shelf utility mining (OSUM) \cite{lan2011discovery} not only considers the utility of patterns, but also utilizes the on-shelf time period related to each pattern. In general, the task of OSUM is to discover the relative high-utility patterns in its selling time periods. OSUM can gain highly useful and valuable information about the underlying relationships in the database, and these discovered patterns are generally more interesting and valuable than frequent patterns. In this field, the first algorithm for OSUM is TP-HOUN \cite{lan2011discovery}, which discovers the on-shelf high-utility itemsets after two phases. After that, several one-phase algorithms, such as FOSHU \cite{fournier2015foshu}, OSUMI \cite{chen2020osumi}, $\rm OSUMI^{+}$ \cite{chen2022shelf}, and OHUQI \cite{chen2021mining}, were proposed to solve the OSUM problem.

Although these algorithms can solve the tasks of OSUM, they still face an important problem in practical applications that they request a user-specific minimum utility (\textit{minutil}) threshold \cite{zhang2021tkus}. For non-professional and ordinary users, setting this \textit{minutil} threshold properly is considered a difficult task before OSUM begins. Since datasets have different densities and different sizes, this \textit{minutil} threshold corresponding to the same number of results will be different. If it is set too high, only a few or no patterns will be discovered. Such a few results may not be enough for the user, since they may be very common. If it is set too low, too many patterns will be found. This situation will require a significant amount of time and memory to mine the overdone results and may even result in confusion. To address the above problem, the issue of top-$k$ pattern mining was proposed, where $k$ is set by the user. In utility mining, the top-$k$ patterns are the $k$ itemsets having the highest utility values, and thus setting $k$ is used instead of the threshold and to directly satisfy the common user's requirements. However, it is still difficult to adopt this technique for finding complete top-$k$ patterns. The key point is that top-$k$ pattern mining algorithm needs to store potential top-$k$ patterns in memory, and it will suffer from performance drawbacks with high memory costs. Besides, it usually requires mining top-$k$ itemsets with negative utility values. This will cause top-$k$ pattern mining tasks to need a long execution time. However, existing algorithms do not satisfy these two points well. It is reflected in algorithms that must keep numerous candidates generated during the mining process in memory and are unable to handle datasets containing items with negative utility.

In this paper, we focus on a generic algorithm for top-$k$ on-shelf utility mining from transaction databases. More precisely, we consider both positive and negative utility values in the database with a value of $k$ specified by users. It means that the $k$ highest on-shelf utility itemsets will be output, and this does not depend on the \textit{minutil} threshold. Then, we proposed a depth-first search algorithm named TOIT for mining Top-$k$ On-shelf hIgh-utility paTterns from a time-period-based database with or without negative utility. Based on the above discussion, the key contributions of TOIT are listed below:

\begin{itemize}
	\item  To solve an important issue about the effectiveness problem of utility mining tasks, TOIT considers both positive and negative utility values. To reduce the consumption of memory and shorten the execution time of calculating, the projection of database and transaction merging techniques are further applied.
	
	\item Two upper-bounds with consideration of on-shelf factor w.r.t. time periods, namely subtree utility and local utility, have been applied in TOIT to prune the search space in an effective way. Unpromising items or itemsets are pruned by these two upper-bounds, so TOIT can discover the top-$k$ results without generating too many candidates.
	
	\item TOIT utilizes the real item utility (RIU) strategy to raise the minimum on-shelf utility value among the top-$k$ high-utility itemsets. This can early prune the unpromising itemsets and narrow the search space, as shown in extensive experimental results.
\end{itemize}

The rest of the paper is organized as follows. The related works on HUPM and OSUM are described in Section \ref{sec:relatedwork}. Then the key preliminaries, definitions, and notations are presented in Section \ref{sec:preliminaries}. The details of our proposed TOIT algorithm are presented in Section \ref{sec:algorithm}. The experimental analysis and detailed results are provided in Section \ref{sec:experiments}. Finally, the conclusion and discussion of our future work are summarized in Section \ref{sec:conclusion}.

\section{Related Work}
\label{sec:relatedwork}

In this section, we review the related work on high utility itemset mining, on-shelf utility mining, and top-$k$ pattern mining.

\subsection{High Utility Itemset Mining}

High utility itemset mining (HUIM) was originally defined by Chan \textit{et al.} \cite{chan2003mining}, for dealing with the limitations of frequency-based itemset mining (FIM). The task of HUIM is to find a complete set of high-utility itemsets (HUIs) whose utility is no less than a minimum user-defined utility threshold. In this task, HUIs generate high utility (e.g., profit) no matter how frequently they show up. However, HUIM is more difficult than FIM for the reason that the utility measurement of itemsets is irregular, i.e., neither monotonic nor anti-monotonic \cite{song2021generalized,gan2019survey}. For example, in FIM, the support of an itemset is always less than or equal to the support of its subset. In HUIM, an itemset's utility may be lower, higher, or equal to the utility of its subsets. Therefore, a downward-closure property based on the Transaction-Weighted Utilization (TWU) \cite{liu2005two} was proposed to solve this problem. This model serves as the upper bound on utility for the itemset or for any sub-itemset in the database. Since then, the Two-Phase \cite{liu2005two}, UP-Growth \cite{tseng2010up}, UP-Growth+ \cite{tseng2012efficient}, BAHUI \cite{song2014bahui} and PB \cite{lan2014efficient} algorithms have been proposed to solve the HUIM task based on the TWU concept. The Two-Phase algorithm \cite{liu2005two} is the first algorithm which applies the TWU model. In the first phase, the algorithm discovers the high TWU itemsets as the candidates for HUIs. In the second phase, the algorithm calculates the actual utility values of the candidates to find out true HUIs by scanning the database. Although this algorithm can finish the task of HUIM, it suffers from the problem of generating too many candidates and scanning the database repeatedly. UP-Growth \cite{tseng2010up} then uses a novel data structure called UP-Tree (Utility Pattern Tree) to store compact information. The candidate itemsets can thus be generated with only two scans of the database. In the first phase, the DLU (Discarding local unpromising items) strategy and DLN (Decreasing local node utilities) strategy are applied to reduce the quantity of candidates and generate potential high utility itemsets (PHUIs) from the global UP-Tree. Although UP-Growth improves the performance of the HUIM, it still has not solved the problems of generating too many candidates and scanning the database frequently. Then EFIM \cite{zida2015efim} proposes an array-based utility counting technique named Fast Utility Count (FAC). Based on the proposed strategies, EFIM can achieve better performance than previous algorithms.

In addition to efficiency, effectiveness is also an important issue in the mining task \cite{gan2019survey}. Many algorithms and models have been proposed over the last few decades to solve the effectiveness problem of utility-oriented pattern mining, including \cite{nguyen2019mining, nguyen2021efficient, hackman2019mining, gan2018survey, gan2018privacy}. In this issue, negative values are commonly seen. Note that there are both positive and negative utility items. Negative items are assumed as the symbol of discounted products sold below cost. Therefore, the problem was redefined as utility mining with both positive and negative values. To address this problem, the HUINIV-Mine algorithm \cite{chu2009efficient} was first proposed. HUINIV-Mine can identify HUIs with negative values in databases, but it generates numerous candidates and consumes a lot of memory. In order to solve this limitation, a single-phase algorithm named FHN \cite{lin2016fhn} was proposed, and it does not generate too many candidates. Then a pattern-growth based algorithm with negative item values, called EHIN \cite{singh2018mining}, was proposed. EHIN employs the projected database technique to reduce memory consumption.

\subsection{On-shelf Utility Mining}

In the actual sale situation, the time of each product on the shelf may be different. This makes traditional HUIM algorithms biased towards products that have been on the shelf for longer periods of time, while ignoring products with higher profits per unit of time. For this limitation, Lan \textit{et al.} \cite{lan2011discovery} first introduced a novel problem called on-shelf utility mining (OSUM), and proposed an algorithm named TP-HOUN. Since TP-HOUN can only handle the database with positive utility, Lan \textit{et al.} \cite{lan2014shelf} proposed a three-scan algorithm named TS-HOUN to discover the on-shelf HUIs with items associated with both positive and negative profit values, although it still has certain drawbacks. It needs to scan the database three times, so its performance in large or dense databases is not satisfactory. Besides, it needs to retain numerous candidates and perform multiple database scans, so it requires a lot of running time and memory space. To address these issues, unlike the previous algorithms, FOSHU \cite{fournier2015foshu} can discover itemsets in a single phase without generating too many candidates, and calculates all time periods at the same time. Thus, it saves time for merging utility-list operations and avoids memory consumption caused by retaining candidate sets. Recently, to deal with sequence data \cite{gan2020fast,wu2021hanp}, Zhang \textit{et al.} \cite{zhang2021shelf} studied the more complicated problem of on-shelf utility mining in sequence data.

\subsection{Top-$k$ Pattern Mining}

The result of traditional HUIM depends on the database and the minimum threshold is set by the user. This mechanism assumes that all users have a good understanding of the database and parameter settings, which is not friendly for users. If the user wants to gain a proper number of results from different databases, the minimum threshold needs to change as the database varies \cite{vo2020mining}. So the issue of top-$k$ pattern mining was proposed, where $k$ is set by the user and the top-$k$ patterns are the $k$ itemsets that have the highest utility values. Two algorithms, called TKU and TKO \cite{tseng2015efficient}, are proposed to discover top-$k$ HUIs. TKU is a two-phase-based algorithm using the TWU model, and it maintains the transaction information and utility of itemsets using a compact tree-based structure called UP-tree. It also proposes five strategies to raise the \textit{minutil}. TKO adopts a different design concept from the former with better performance. More precisely, TKO utilizes a list-based structure called utility-list \cite{liu2012mining}. Based on this data structure, TKO supplied novel space pruning strategies to mine top-$k$ HUIs in only one phase. Compared to TKU, the REPT algorithm \cite{ryang2015top} was presented with its special strategies. It discovers the results from two database scans. In its first scan, REPT adopts two strategies named PUD (Pre-evaluation with Utility Descending order) and RIU (Real Item Utilities) to increase an initial threshold value. In its second scan, REPT uses another strategy called RSD (support descending order) to raise the threshold again. However, REPT essentially is a two-phase algorithm and its efficiency still has room to be improved. Since then, an efficient algorithm named $k$HMC \cite{duong2016efficient} was designed to solve the task of efficiency. $k$HMC is a single-phase algorithm, which uses three strategies to increase its \textit{minutil} and thus reduce the memory cost. More precisely, there are two new pruning strategies in $k$HMC to reduce the search space. On the one hand, to avoid the cost of merging operations between the utility-list, $k$HMC proposes a novel co-occurrence pruning technique named EUCPT. On the other hand, another strategy named TEP is used to get a novel upper-bound on the utility of itemsets. The result of experiments shows that $k$HMC outperforms the TKO and REPT algorithms. Inspire by $k$HMC, KOSHU \cite{dam2017efficient} utilizes three strategies to reduce the search space. The first strategy is EMPRP (efficient estimated co-occurrence maximum period rate pruning), the second is PUP (period utility pruning), and the last one is CE2P (concurrence existing of a pair of 2-itemset pruning). KOSHU also supplies two \textit{minutil} raising strategies to improve efficiency. To the best of our knowledge, KOSHU is the most efficient algorithm in top-$k$ pattern mining, and we will use it as our baseline algorithm in this paper.

\section{Preliminaries and Problem Statement}
\label{sec:preliminaries}

In this section, we first introduce the main concepts of top-$k$ on-shelf utility mining from transaction databases. We then show the related notations of these concepts. The detailed definitions are shown below.

\begin{table}[h]
	\caption{A sample transaction database}
	\label{table:database}
	\centering
	\begin{tabular}{|c|l|c|}
		\hline
		\textbf{Tid}	&	\textbf{Transaction}	&	\textbf{Period} \\ \hline
		$T_1$	&	$(a,1)$ $(b,2)$ $(d,4)$ $(e,1)$	&	1\\ \hline
		$T_2$	&	$(b,1)$ $(c,2)$ $(d,12)$	&	0\\ \hline
		$T_3$	&	$(a,3)$ $(d,10)$	&	1\\ \hline
		$T_4$	&	$(a,1)$ $(e,1)$	&	2\\ \hline
		$T_5$	&	$(b,1)$ $(c,2)$ $(d,12)$ $(e,1)$	&	2\\ \hline
		$T_6$	&	$(b,1)$ $(c,1)$ $(e,2)$	&	2\\ \hline
		$T_7$	&	$(a,2)$	&	0\\ \hline
		$T_8$	&	$(b,1)$ $(c,1)$ $(d,8)$	&	1\\ \hline
	\end{tabular}
\end{table}

\begin{definition}(Database)
	\rm Assume a database $\mathcal{D}$ = \{$T_1$, $T_2$, $\cdots$, $T_n$\} is the running example database for on-shelf utility mining, and $I$ = \{$i_1$, $i_2$, $i_3$, $\cdots$, $i_n$\} is a set of $n$ items which appears in $\mathcal{D}$. Each transaction $T \in D$ has its own transaction identity named $tid$. The $tid$ of $T_c$ $\in D$ is $c$. For each item $i \in T_c$, there is an internal utility (a positive value) and denoted as $q(i, T_c)$; and an external utility (either positive or negative value) denoted as $p(i)$. One or more items can form an itemset which is defined as $X$ $\subseteq$ $I$. Because $\mathcal{D}$ is an on-shelf transaction database, each transaction, such as $T_c$, has its own time period, which is called $pt(T_c)$. The collection of all time periods appearing in $\mathcal{D}$ is defined as \textit{PE}.
\end{definition}

Take Table \ref{table:database} as an example database $\mathcal{D}$. It contains eight transactions \{$T_1$, $T_2$, $\cdots$, $T_8$\} and three time period \textit{PE} = \{0, 1, 2\}. The transaction $T_3$ in time period 1 contains two items, $a$ and $d$, respectively. The internal utility of $a$ is 3 and the internal utility of $d$ is 10. The utilities of these two items are, respectively, 5 and 3, which are indicated in Table \ref{table:profit}.

\begin{table}[!htbp]
	\caption{External utility values}
	\label{table:profit}
	\centering
	\begin{tabular}{|c|c|c|c|c|c|}
		\hline
		\textbf{Item}	&	$a$	&	$b$	&	$c$	&	$d$	&	$e$	\\	\hline
		\textbf{Utility (\$)}	&	5	&	-3	&	-2	&	3	&	10	\\	\hline
	\end{tabular}
\end{table}

\begin{definition}{(Utility of item and itemset)}
	\rm The utility of an item $i$ and an itemset $X$ in the database $\mathcal{D}$ are respectively defined as $u(i)$ and $u(X)$. In transaction $T_c$, the utility of $i$ is $u(i, T_c)$ = $p(i) \times q(i, T_c)$ where $i \in T_c$; the utility of $X$ in $T_c$ is defined as $u(X, T_c)$. The relationship between $u(i, T_c)$ and $u(X, T_c)$ is $u(X,T_c)$ = $\sum_{i \in X \wedge X \subseteq T_c} u(i,T_c)$. $u(X)$ = $\sum_{X \subseteq T_c \wedge T_c \in D}u(X,T_c)$.
\end{definition}

The utility of the item $a$ in $T_1$ is $u(a, T_1)$ = 1 $\times $ \$5 = \$5. The utility of the itemset \{$c, e$\} in $T_6$ is $u(\{c, e\}, T_6)$ = $u(c, T_6)$ + $u(e, T_6)$ = 1 $\times$ \$(-2) + 2 $\times $ \$10 = \$18. $u(\{c, e\})$ = $u(\{c, e\}, T_6)$ + $u(\{c, e\}, T_5)$ = \$18 + \$6 = \$24.

\begin{definition}(Time period)
	\rm Each transaction has only one time period, but an itemset $X \subseteq I$ may have zero or more time period. Definition of time period of itemset $X \subseteq I$ is $pi(X)$ = $\{pt(T_c) | T_c \in D \wedge X \subseteq Tc\}$.
\end{definition}

\begin{definition}(Utility of a transaction)
	\rm Assume that $T_c$ is a transaction in $\mathcal{D}$. The utility of $T_c$, which is equal to the sum utility of all items in $T_c$, is denoted as $TU(T_c)$ = $ \sum_{i \in T_c} u(i,T_c)$. The utility of all positive items in $T_c$ is defined as $PTU(T_c)$ = $\sum_{i \in T_c \wedge p(i) > 0} u(i,T_c).$
\end{definition}

The time period of $\{b, e\}$ is $pi(\{b, e\})$ = \{1, 2\}. Consider transaction $T_2$, the utility of $T_2$ is $TU(T_2)$ = $u(b, T_2)$ + $u(c, T_2)$ + $u(d, T_2)$ = 1 $\times $ \$(-3) + 2  $\times $ \$(-2) + 3 $\times $ \$12 = \$24. The utility of all positive items in $T_2$ is $PTU(T_2)$ = $u(d, T_2)$ = 3 $ \times $ \$12 = \$36.

\begin{definition}(Utility of an itemset in a time period)
	\rm We define the utility of an itemset $X$ in a time period $h \in pi(X)$ as $u(X,h)$ = $\sum_{T_c \in D \wedge h = pt(T_c) \wedge X \subseteq T_c}$ $u(X,T_c)$.
\end{definition}

For example, we can calculate the utility of itemset $\{c, e\}$ in time period $2$ by $u(\{c, e\}, 2)$ = $u(\{c, e\},T_5)$ + $u(\{c, e\},T_6)$ = \$6 + \$18 = \$24.

\begin{definition}(Sum utility and Relative utility)
	\rm Define $to(X)$ as the sum utility of itemset $X$ which has the same time period in all the transactions, the formula is $to(X)$ = $\sum_{h \in pi(X) \wedge T_c \in D \wedge h= pt(T_c)} TU(T_c)$. Then we define the relative utility ($ru$) of $X$ as the ratio of the utility of $X$ in the current time period and the sum utility of $X$. The formula that calculate the relative utility ($ru$) of $X$ is $ru(X)$ = $u(X)$/$to(X)$.
\end{definition}

The time period of $\{b, e\}$ is $\{1, 2\}$. Therefore, $to(\{b, e\})$ = $TU(T_1)$ + $TU(T_3)$ + $TU(T_4)$ + $TU(T_5)$ + $TU(T_6)$ +$ TU(T_8)$ = \$21 + \$45 + \$15 + \$39 + \$15 + \$19 = \$154. And the utility of $\{b, e\}$ is $u(\{b, e\})$ = $u(\{b, e\}, T_1)$ + $u(\{b, e\}, T_5)$ + $u(\{b, e\}, T_6)$ = \$4 + \$7 + \$17 = \$28. It can be obtained from the above result that $ru(\{b, e\})$ = $u(\{b, e\})$ / $to(\{b, e\})$ = \$28/\$154 = 0.181.

In top-$k$ pattern mining, we define the internal utility as \textit{interutil}, which is equal to zero at the beginning and raised in the follow-up process. At the end of the process, the value of \textit{interutil} may equal to the relative utility of the $k$-th itemset if the number of patterns is no less than $k$.

\begin{definition}(On-shelf high utility itemset)
	\rm We define the number of patterns which will be discovered as $k$. Then, an on-shelf high utility itemset is considered as the top $k$ patterns according to descending order of relative utility.
\end{definition}

\begin{definition}(Transaction weighted utilization)
	\rm  The transaction weighted utilization, called $TWU$, of an itemset $X$ is defined as $TWU(X)$ = $\sum_{T_c \in D \wedge X \subseteq T_c} PTU(T_c)$. The $TWU$ of itemset $X$ in a specific time period $h$ is denoted as $TWU(X,h)$ = $\sum_{T_c \in D \wedge X \subseteq T_c \wedge pt(T_c) = h}$ $PTU(T_c)$.
\end{definition}

The $TWU$ of $\{b, c\}$ is $TWU(\{b, c\})$ = $PTU(T_2)$ + $PTU(T_5)$ + $PTU(T_6)$ + $PTU(T_8)$ = \$36 + \$46 + \$20 = \$102. The $TWU$ of $\{c, e\}$ in time period $2$ is $TWU(\{c, e\}, 2)$ = $PTU(T_5)$ + $PTU(T_6)$ = \$46 + \$20 = \$66.

\begin{definition}(Relative utility of an itemset in a time period)
	\rm The sum utility of all transactions in the same time period $h$ is denoted as $pto(h)$ = $ \sum_{T_c \in D \wedge h = pt(T_c)} TU(T_c)$. We then define the relative utility of itemset $X$ in time period $h$ as $ru(X,h)$ = $u(X,h)$ / $pto(h)$.
\end{definition}

To calculate the relative utility of $\{b, e\}$ in time period $1$, we compute the $pto(1)$ = $TU(T_1)$ + $TU(T_3)$ + $TU(T_8)$ = \$21 + \$45 + \$19 = \$85, and $u(\{b, e\}, 1)$ = $u(\{b, e\}, T_1)$ = \$4. Then the relative utility of $\{b, e\}$ is $ru(\{b, e\}, 1)$ = $u(\{b, e\}, 1)$ / $pto(1)$ = \$4 / \$85 = 0.04.

\begin{property}
	\rm The upper-bound of utility of itemset $X$ in time period $h$ is $TWU(X, h)$, i.e., $TW U(X, h)$ $\geq$ $u(X, h)$.
\end{property}

\begin{property}
	\rm If no time period $h$ for an itemset $X$ satisfied $TWU(X, h)$/$pto(h)$ $\geq $ \textit{interutil}, we can consider that $X$ and its supersets are not on-shelf high utility itemsets.
\end{property}

\begin{definition}(Remaining utility)
	\rm Assume the symbol $\succ$ is the order of items from any itemset $I$. $\succ$ stipulates that the items should be arranged in $TWU$ ascending and positive items in front of negative items order. Then for an itemset $X$ in a transaction $T_c$, we define that its positive remaining utility as $pre(X, T_{c})$ =  $\sum_{i \in T_{c} \wedge i \succ x \forall x \in X \wedge p(i)>0} u(i, T_{c})$, we define its remaining utility as  $re(X, T_{c})$ = $\sum_{i \in T_{c} \wedge i \succ x \forall x \in X} u(i, T_{c})$. For an itemset $X$ in a time period $h$, its positive remaining utility is defined as $pre(X, h)$ = $\sum_{T_c \in D \wedge pt(T_c) =  h \wedge X \subseteq T_c} pre(X, T_{c})$, and its remaining utility is defined as $re(X, h)$ = $\sum_{T_c \in D \wedge pt(T_c) = h \wedge X \subseteq T_c}$ $re(X, T_{c})$.
\end{definition}

For example, we sort the items by $\succ$ order $\{d \succ b \succ c \succ e \succ a\}$. The positive remaining utility of $\{a, e\}$ in time period 1 can be calculated as $pre(\{a, e\}, 1)$ = $pre(\{a, e\}$, $T_{1})$ = 4 $\times $ \$3 = \$12. The remaining utility of $\{b, e\}$ in time period $2$ can be calculated as $re(\{b, e\}, 2)$ = $re(\{b, e\}$, $T_{5})$ + $re(\{b, e\}$, $T_{6})$ = 12 $\times$ \$3 + 0 = \$36.

\begin{definition}(Utility-list \cite{liu2012mining})
	\rm Utility-list is a data structure which stores the information of itemsets. There are three attributes, (\textit{tid}, \textit{iutil}, \textit{rutil}), in the utility list. For itemset $X$, $tid$ is the ID number of the transaction, \textit{iutil} is the utility of $X$ in transaction $T_{tid}$ and \textit{rutil} stands for the positive remaining utility of $X$ in $T_{tid}$.
\end{definition}

\begin{property}
The remaining utility of itemset $X$ in time period $h$ has an upper-bound that is $reu(X,h)$ = $u(X,h)$ + $re(X,h)$.
\end{property}

For example, $\{b, e\}$ in time period $2$ has $reu(\{b, e\}, 2)$ = $u(\{b, e\}, 2)$ + $ re(\{b, e\}, 2)$ = (\$7 + \$17) + \$36 = \$60.

\begin{property}
	\rm If no time period $h$ for an itemset $X$ satisfied $reu(X,h)$ / $pto(X,h)$ $\geq $ \textit{interutil}, $X$ and all its extensions can be considered as low on-shelf utility itemsets.
\end{property}

\begin{property}
	\rm Let $X$ and $Y$ be two itemsets with $X \subset Y$. The relationship of $TWU$ of these itemsets in the time period $h$ is $TWU(X, h)$ $\geq$ $TWU(Y, h)$.
\end{property}

\begin{property}
	\rm The relative utility of $X$ in the time period $h$ has an upper-bound that is defined as the ratio of the $TWU$ of an itemset $X$ in a certain period $h$ divided by the total utility in that time period $h$, i.e. $TWU(X,h)$/$pto(h)$ $\geq ru(X,h)$.
\end{property}

\section{The TOIT Algorithm}
\label{sec:algorithm}

The relevant algorithms proposed at present still need performance improvement. Therefore, in this section, we propose a single-phase on-shelf high utility itemset mining algorithm named TOIT, which can handle items with positive or negative utility. Compared with the state-of-the-art KOSHU algorithm, TOIT avoids the cost of utility-list construction operations.

\subsection{Search Space}

\begin{figure}
	\centering
	\includegraphics[scale=0.5]{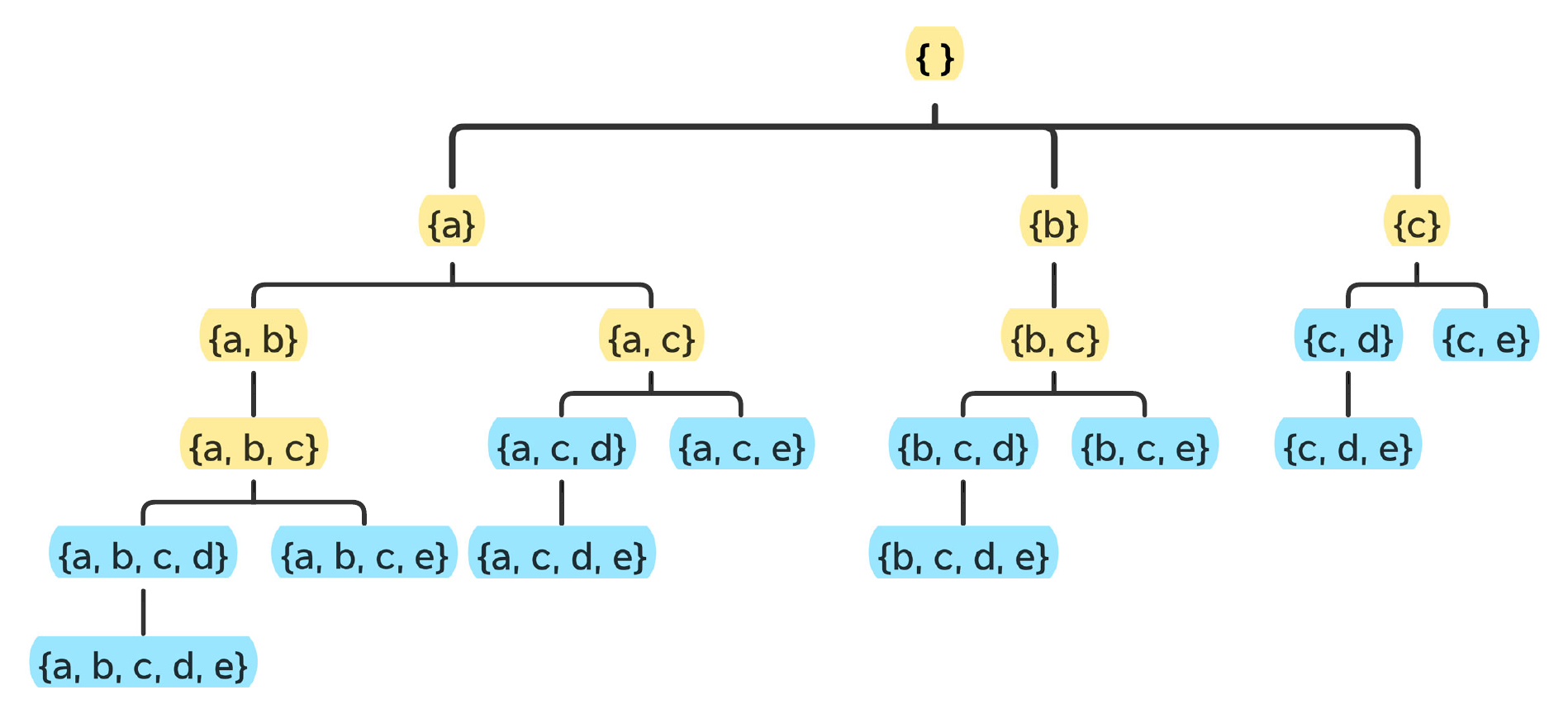}
	\caption{A set-enumeration tree for $I$ = \{$a$, $b$, $c$, $d$, $e$\}}
	\label{fig:tree}
\end{figure}

The search space of TOIT is presented as a \textit{set-enumeration tree} \cite{liu2012mining,lin2016fhn}. TOIT begins to do a depth-first search at the root of the tree with an empty itemset. While doing a depth-first search, the current itemset adds an item from the extension itemset to itself to form a new itemset. For handling both the positive and negative items, the rule of expansion is shown as follows. According to the ascending order of TWU, TOIT first selects the items with positive effects, and then selects the negative item if there are no positive items in the extension itemset of the current itemset. In Fig. \ref{fig:tree}, we assume the $\{a, b, c\}$ items with a positive utility and the $\{d, e\}$ items with a negative utility\footnote{This assumption is not based on Table \ref{table:profit}}. The nodes with the yellow color are expanded by positive items, and the nodes with the blue color are expanded by negative items.

\begin{definition}(Expansion)
	\rm We define the expansion items of itemset $\alpha$ as $E(\alpha)$ = $\{z |$ $z \in I \wedge z$ $\succ x$ $\wedge \forall x$ $\in \alpha\}$. Assume $Z$ is one of the expansion of $\alpha$. In the set-enumeration tree it is presented as a subtree of $\alpha$, and is denoted as $Z$ = $\alpha$ $\cup W$, where $W \subseteq E(\alpha)$ and $W \neq \emptyset$. Specially if $Z$ is a one-item extension of $\alpha$, we define it as $Z$ = $\alpha \cup \{z\}$, where $z \in E(\alpha)$.
\end{definition}

\subsection{Database Projection}

A database projection strategy is applied in the TOIT algorithm, which can reduce the resource consumption while calculating the upper-bound of the current itemset by scanning the current database.

\begin{definition}(Projected database)
	\rm The projected database \cite{pei2004mining,lan2014eff,zida2015efim} consists of a set of  projected transaction. Given an itemset $\alpha$, we denote the projected transaction of a transaction $T$ as $\alpha-T$ = $\{i \mid i \in T \wedge i \in E(\alpha)\}$. Given an itemset $\alpha$, we denote the projected database of a database $\mathcal{D}$ as $\alpha|_D$ = $\{\alpha-T \mid T \in D \wedge \alpha-T \neq \emptyset\}$.
\end{definition}

Following the construction of the projected database based on the itemset $\alpha$, those items ($x \notin E(\alpha)$) that are not included in the expansions of $\alpha$ had to be removed. Because it only considers the promising items ($y \in E(\alpha) $), this operation improved the depth-first search efficiency. In other words, the projected database based on $\alpha$ only considers the situation in the subtree of $\alpha$. Before using a projected database strategy, two pre-operations must be completed. First, we need to store the transactions in different time periods according to the $\succ$ total order, respectively. Second, we need to set a positive and a negative offset pointer on the corresponding original transaction to build the \textit{pseudo-projected database}.

\subsection{Transaction Merging}

In the case of on-shelf utility mining, all transactions in the database are divided into several time periods. In each time period, there may be multiple identical transactions that have the same time period and items (but no need for the same internal utility of each item). To reduce the cost of scanning these identical transactions during the recursive search, we apply a transaction merging strategy to merge the identical transactions in each projected database. This strategy can reduce the size of each projected database and does not affect the final result.

\begin{definition}(Transaction merging \rm \cite{zida2015efim})
	\rm To merge a set of identical transactions ($T_{k1}$, $T_{k2}$, $\cdots$, $T_{kn}$) in a database $\mathcal{D}$, we know that $T_{k1}$ = $T_{k2}$ = $\cdots$ = $T_{kn}$, and then we can build a new transaction $T_{M}$ in the same period $h$ to replace them. Then we define the time period of $T_M$ as $h$ and the quantity of each item $i \in T_M$ as $q(i, T_{M})$ = $\sum_{k=1 \ldots n}$$q(i, T k_{n})$, respectively.
\end{definition}

\begin{definition}(Projected transaction merging \rm \cite{zida2015efim})
	\rm Consider a set of identical transactions ($T_{k1}$, $T_{k2}$, $\cdots$, $T_{kn}$) in projected database $\alpha|_D$, to merge them we can also build a new transaction $T_{M}$ to replace them. Then we define the time period of $T_M$ as $h$ and  the quantity of each item $i \in T_M$ as $q(i, T_{M})$ = $\sum_{k=1 \ldots n} q(i, T k_{n})$.
\end{definition}

There is still a problem here: how to find the same transactions efficiently? We can compare each transaction in the same time period to discover the sets of identical transactions. It is a viable method, but it takes $O(n^2)$ time to complete this task for every $n$ transactions in a database in the worst-case scenario. To efficiently solve this problem, we perform a pre-operation that sorts transactions for each time period using a new total order $\succ_T$, and this needs $O(nlog_n)$ time. The complexity of discovering identical transactions in a database can be reduced to $O(n) $ with this performance. In comparison to the worst-case scenario, which takes $O(n^2) $ time, TOIT's $O(nlog_n) $ solution is clearly more efficient.

\begin{definition}(Transaction sorting)
	\rm $\succ_T$ is a symbol, which means that we do a lexicographical order of the transactions backwards. According to the following four steps, we can judge the relationship between two transactions in the same time period.
	
	\begin{itemize}
		\item We compare the last item of two compared transactions to find out the larger item in the lexicographical order. The transaction that has the larger item is the larger transaction.
		
		\item If there are equal items in the first step, the previous items of the two compared transactions are used to compare until we find out the larger one. The transaction having the larger item is the larger transaction.
		
		\item If the comparing result is always equal and one of the transactions has no additional items to continue comparing, the transaction which has remaining items (the longer transaction) is considered to be the larger transaction.
		
		\item If the length of two transactions is equal, the two transactions are considered equal.
	\end{itemize}
\end{definition}

\begin{property}
	\rm After sorting the transaction according to the $\succ_T$ order, the identical transactions in the same time period are lined up together. It works on both the original database and the projected database.
\end{property}

\subsection{Subtree Utility and Local Utility}
\label{subsec:SuLu}

In order to narrow the search space and reduce the memory consumption of pattern mining, two strategies based on upper-bounds named subtree utility and local utility are applied in TOIT. By doing so, we find that the efficiency of the mining process can be significantly improved. The detailed introduction of these two strategies is listed as follows.

\begin{definition}(Subtree utility \rm \cite{zida2015efim})
	\rm We assume that $\alpha$ is an itemset, $z$ is one of the extension items of $\alpha$ ($z \in E(\alpha)$) and $h$ is a time period. We define $su(\alpha, z, h)$ to represent the subtree utility of $z$ based on $\alpha$ in the time period $h$. Then we have $su(\alpha, z, h)$ = $\sum_{T \in g(\alpha \cup\{z\}) \wedge pt(T) = h}[u(\alpha, T)$ + $u(z, T)$ + $\sum_{i \in T \wedge i \in E(\alpha \cup\{z\})} u(i, T)]$ \cite{zida2015efim}.
\end{definition}

\begin{property}
	\rm In the time period $h$, the formula $su(\alpha, z, h) \geq u(\alpha \cup\{z\}, h)$ always exists for an itemset $\alpha$ and its extension item $z$.  In addition, we suppose that $Z$ is the extension of $\alpha \cup \{z\}$ and inequality $su(\alpha, z, h) \geq u(Z, h)$ is also satisfied in the case of any $Z$ in $h$. Then we have $su(\alpha, z, h)$/$pto(h)$ $\geq$ $u(\alpha \cup\{z\}, h)$/$pto(h)$ and $su(\alpha, z, h)$/$pto(h)$ $\geq$ $u(Z, h)$/$pto(h)$.
\end{property}

\begin{theorem}(Using subtree utility to prune the unpromising subtree)
	\label{theorem:su}
	\rm We assume that $\alpha$ is an itemset, $z$ is an extension item of $\alpha$ ($z \in E(\alpha)$), and $h$ is a time period. We consider that $\alpha \cup \{z\}$ and all its extensions are low-utility if there is no such an $h$ satisfy $su(\alpha, z, h)/pto(h)$ $\geq $ \textit{interutil}. If this inequality is hold on the set-enumeration tree, the subtree of $\alpha \cup\{z\}$ is low-utility, and it can be trimmed.
\end{theorem}

\begin{definition}(Local utility \rm \cite{zida2015efim})
	\rm We assume that $\alpha$ is an itemset, $z$ is an extension item of $\alpha$ ($z \in E(\alpha)$), and $h$ is a time period. Then the local utility of $\alpha$ and $z$ in $h$ is denoted as $lu(\alpha, z, h)$ = $\sum_{T \in g(\alpha \cup\{z\}) \wedge pt(T)=h }$$[u(\alpha, T)$ + $ re(\alpha, T)]$ \cite{zida2015efim}.
\end{definition}

\begin{property}
	\rm For itemset $\alpha$ and its expand item $z$ in time period $h$, the inequality $lu(\alpha, z, h)$ $\geq$ $u(\alpha \cup\{z\}, h)$ is always holding.
\end{property}

\begin{theorem}(Using local utility to prune the unpromising item)
	\label{theorem:lu}
	\rm We assume that $\alpha$ is an itemset, $z$ is an extension item of $\alpha$ ($z \in E(\alpha)$), and $h$ is a time period. We realize that any extensions of $\alpha$ that contain $z$ have a low utility if there is no such a $h$ satisfy $lu(\alpha, z, h)$/$pto(h)$ $\geq $ \textit{interutil}. This situation appears on the set-enumeration tree is that the $z$ can be pruned in the subtree of $\alpha$.
\end{theorem}

TOIT uses these theorems mentioned above to narrow the search space. First, it applies Theorem \ref{theorem:su} to cut off the unpromising subtree of the current itemset. Second, it uses Theorem \ref{theorem:lu} to remove the item which is unpromising in the remaining subtrees.

\begin{property}(Relationship of upper-bound)
	\label{property:upper-bound}
	\rm We assume that $\alpha$ is an itemset, $z$ is an extension item of $\alpha$ ($z \in E(\alpha)$), $Y$ is an itemset  $Y$ = $\alpha \cup \{z\}$ and $h$ is a time period. The relationship of upper-bound is $TWU(Y, h)$ $\geq$ $lu(\alpha, z, h)$ $\geq$ $reu(Y, h) $ = $ su(\alpha, z, h)$ \cite{zida2015efim}.
\end{property}

\begin{definition}(Primary, secondary, and negative itemset \rm \cite{zida2015efim})
	\rm	The \textit{Primary} itemset of itemset $\alpha$ is defined as \textit{Primary}($\alpha$) = $\{i \mid i \in $ \textit{Secondary}($\alpha) \wedge (\exists h)$ $(su(\alpha, i, h)$/$pto(h))$ $\geq$ \textit{interutil}\}. The \textit{Secondary} itemset of itemset $\alpha$ is defined as \textit{Secondary}($\alpha$) = $\{i \mid i \in I \wedge q(i) \geq 0 \wedge (\exists h)$ $(lu(\alpha, i, h)$/$pto(h))$ $\geq$ \textit{interutil}\}. Because $lu(\alpha, z)$ $\geq$ $su(\alpha, z)$ holds, the connection between these two itemsets is \textit{Primary}($\alpha)$ $\subseteq$ \textit{Secondary}($\alpha)$.	To handle the negative values, we define the \textit{Negative} itemset as \textit{NegativeItems}($\beta)$ = $\{n \mid q(n) < 0 \wedge (\exists h)$ $(su(\beta, n, h)$ / $pto(h) >$ \textit{minutil})\}.
\end{definition}

According to the relationship mentioned in Property \ref{property:upper-bound}, we could conclude that a local utility has a tighter upper-bound than TWU. In other words, local utility is more effective in pruning the search space. Besides, subtree utility is also worthy of attention. Although $su$ is equal to $reu$ in value, the advantage of subtree utility is that it can be calculated in the process of searching $\alpha$. The value of $reu$, on the other hand, is calculated during the search for $Y$. In other words, because subtree utility is calculated earlier than $reu$, it can cut off the meaningless extensions of the current itemset to improve the search efficiency.

\subsection{Upper-Bounds in Arrays}

In this section, we introduce a special data structure called the on-shelf utility two-dimensional array (OUTA) to efficiently compute the subtree utility and local utility (described in subsection \ref{subsec:SuLu}). Unlike the previous algorithm, using OUTA to compute the upper-bound can avoid the operation of utility-list construction. Therefore, the performance of TOIT is better than other algorithms both in time consumption and memory cost.

\begin{definition}(On-shelf utility two-dimensional array \rm \cite{chen2020osumi})
	\rm We define that $tp_n$ is the size of \textit{PE} ($tp_n = |PE|$) and $i_n$ is the size of $I$ ($i_n$ = $|I|$), where \textit{PE} and $I$ are the same in $\mathcal{D}$. We defined OUTA as $U$ with $tp_n$ rows and $i_n$ columns. We can store the value of each item in $U$. Then, $U[h][z]$ represents the value of item $z$ in a time period $h$. When we initialize OUTA, we set all the values to 0.
\end{definition}

\textbf{Calculating TWU by OUTA} After initializing OUTA, we calculate the TWU value of each item $z$ w.r.t. each time period $h$ by formulate $U[h][z]$ = $U[h][z]$ + $TU(T)$, where $T$ is the transactions containing $z$ ($z \in T$). After a scan of the database to calculate the TWU of each item $z \in I$, $U[h][k]$ and $TWU(k, h)$ are equal in value.

\textbf{Calculating subtree utility by OUTA} First, we initialize OUTA to calculate the subtree utility of the current itemset $\alpha$. Second, we calculate the subtree utility of each item $z$ ($z \in T \cap E(\alpha)$) in transaction $T$ w.r.t. each time period $h$, by formulating $U[h][z]$ = $U[h][z]$ + $u(\alpha, T)$ + $ u(z, T)$ + $\sum_{i \in T \wedge i \succ z} u(i, T)$. After a scan of the database to calculate the subtree utility of each item $z \in E(\alpha)$, $U[h][k]$ and $su(\alpha, k, h)$ are equal in value.

\textbf{Calculating local utility by OUTA} First, we initialize OUTA to compute the local utility of the current itemset $\alpha$. Second, we calculate the local utility of each item $z$ ($z \in T \cap E(\alpha)$) in transaction $T$ w.r.t. each time period $h$, by formulating $U[h][z]$ = $U[h][z]$ + $u(\alpha, T)$ + $re(\alpha, T)$. Following a database search to determine the local utility of each item $z \in E(\alpha)$, $U[h][k]$ and $lu(\alpha, k, h)$ are equal in value.

\subsection{Mining Strategies}

\textbf{Real one-item relative utilities raising strategy, RIRU strategy for short}. To discover the top-$k$ itemsets, we apply the RIRU strategy to raise the \textit{interutil}. First, we calculate the relative utility $ru$ of each item $i$ ($i$ $\in$ $I$). We assume that $R$ = \{$ru_1$, $\cdots$, $ru_n$\} ($n$ = $|I|$). If there are no less than $k$ $ru$ elements with a value greater than 0 in $R$, then the value of \textit{interutil} is set to the highest $k$-th $ru$. If there are less than $k$ $ru$ elements with a value greater than 0 in $R$, the value of \textit{interutil} is set to 0. Because the itemset of a single item is a subset of all itemsets, the top-$k$ relative itemsets of a single item are not more than the real top-$k$ relative itemsets. In other words, the relative utility of the $k$-th itemset of a single item, which is set to \textit{interutil}, is no more than the relative utility of the real $k$-th itemset.

\textbf{Dealing with negative items}. In addition to the positive utility items in the database, there are also negative utility items. Therefore, we handle this problem by using a method that extends negative items on the basis of positive itemsets. In HUIM, extending negative items on an itemset will reduce the utility of the extension itemset. But this rule does not apply in the high utility itemset mining situation. The assessment of relative utility is neither monotonic nor anti-monotonic. While doing the searching process, if an itmeset $\alpha$ has a subtree utility $su(\alpha, n, h)$ that is no less than the \textit{interutil} in at least one time period $h$, it will be extended by a negative item $n$ until it does not meet this condition. We express this by the formula $(\exists h)(su(\beta, n, h)$ / $pto(h)$ $>$ \textit{interutil}).

\subsection{Proposed TOIT Algorithm}

In this section, we will introduce some pseudo-codes for the TOIT algorithm, which utilizes several novel and efficient strategies.

% TOIT Algorithm
\begin{algorithm}[ht]
	\caption{The TOIT algorithm}
	\label{Algorithm:TOIT}
	\KwIn{\textit{D}: one on-shelf transaction database, \textit{k}: the number of patterns which is pre-defined by the user.}
	\KwOut{\textit{OSUMs}: a set of itemsets with on-shelf high-utility. }
	\BlankLine
	
	initial $\alpha$ as $ \emptyset$;
	
	compute $\textit{RIU}(k)$;	\tcp{calculate and raise the \textit{interutil}}
	\For{\rm each time period $h \in D$ }{
		\For{\rm each item $i \in I$}{
			calculate $lu(\alpha, i, h)$ using OUTA\;
		}
		
	}
	\textit{Secondary}($\alpha$) = $\{i \mid i \in I \wedge q(i) \geq 0 \wedge (\exists h)(lu(\alpha, i, h)/pto(h))$ $\geq $ \textit{interutil}\}\;
	sort \textit{Secondary}($\alpha$) by total order based on TWU value increasing, which is $\succ$\;
	
	\For{\rm each time period $h \in D$}{
		remove $i \notin$ \textit{Secondary}($\alpha$) from transactions, and then delete empty transactions\;
		sort the transactions in time period $h$ according to $\succ_{T}$ \;
	}
	\For{\rm each time period $h \in D$}{
		calculate the $su(\alpha, i, h)$ of each item $i \in $ \textit{Secondary}($\alpha$) using an OUTA by scanning all transactions in $h$ \;
	}
	compute \textit{Primary}($\alpha$)  = $\{i \mid i \in$ \textit{Secondary}($\alpha$) $\wedge$  $(\exists h)$ $(su(\alpha, i, h)$ / $pto(h))$ $\geq$ \textit{interutil}\}\;
	
	call \textit{Search}$(\alpha$, $\mathcal{D}$, \textit{Primary}($\alpha$), \textit{Secondary}($\alpha$), $k$)\;
\end{algorithm}

The beginning part of the program (cf. Algorithm \ref{Algorithm:TOIT}) requires the input of an on-shelf transaction database and the user-specified pattern amount. At the beginning, it initializes the empty database. Then, it calls the RIU function to raise the value of \textit{interutil}. After that, it searches each time period to compute the local utility using OUTA. While the $\alpha$ is empty ($\alpha$ = $\emptyset$), the value of $lu(\alpha, i, h)$ is equal to TWU of item $i$ in time period $h$. Then it builds the \textit{Secondary}($\alpha$) itemset by the item $i$ that meets the formula $lu(\alpha, i, h)$ /$ pto(h)$ $\geq$ \textit{interutil} at least in one time period $h$. It sorts the items in the \textit{Secondary}($\alpha$) itemset in the order of increasing relative TWU values. It traverses the transactions in each $h$ and removes the items which \textit{Secondary}($\alpha$) does not contain. If all items in the transaction have been removed, this transaction will be deleted. Then, the algorithm sorts the rest of the transactions in each time period by $\succ_T$. TOIT does the scans according to each time period to compute the subtree utility of each extension item by using an OUTA. A positive item $i$ will be added into \textit{Primary}($\alpha$) if it finds a time period $h$ that satisfies $su(\alpha, i, h)$/$pto(h)) \geq $ \textit{minutil}. Finally, TOIT invokes the recursive function \textit{Search} to do a depth-first search beginning with the current itemset $\alpha$. The OSUMs, output by the TOIT algorithm, is a $k$-length set which contains the top-$k$ on-shelf utility itemsets.

% Search Algorithm
\begin{algorithm}[h]
	\caption{The Search procedure}
	\label{Algorithm:Search}
	\KwIn{$\alpha$: a prefix itemset, $\alpha$-$\mathcal{D}$: the projected database based on $\alpha$, \textit{Primary}($\alpha$): a set of primary items of $\alpha$, \textit{Secondary}($\alpha$): a set of secondary items of $\alpha$, $k$: the amount of pattern user need.}
	\KwOut{a set of itemsets which are supersets of $\alpha$ with on-shelf high utility.}
	\BlankLine
	
	\For{\rm each item $i \in $ \textit{Primary}($\alpha)$}{
		set a new super-itemset $\beta$ = $\alpha$ $\cup\{i\}$\;
		compute $u(\beta)$ and create the projected database $\beta$-$\mathcal{D}$ by scanning $\alpha$-$\mathcal{D}$\;
		compute $ru$ of $\beta$\;
		\If{$ru$ $\geq$ \textit{interutil}}{
			add ItemsetToTopK($\beta$)\;
		}
		
		\textit{NegativeItems}($\beta$) = $\{n \mid q(n) < 0$ $\wedge$ $ (\exists h)$ $(su(\beta, n, h)$ / $pto(h) >$ \textit{interutil}) \}\;
		call \textit{NegativeSearch}$(\beta$, $\beta$-$\mathcal{D}$, \textit{NegativeItems}($\beta$), $k$)\;
		
		\For{\rm each time period $h$ in $\beta$-$\mathcal{D}$}{
			\For{\rm  each item $z \in $ \textit{Secondary}($\alpha)$}{
				compute $su(\beta, z, h)$ and $lu(\beta, z, h)$ by using two OUTAs respectively\;
			}			
		}
	
		compute \textit{Primary}($\beta$)  = $\{z \in $ \textit{Secondary}$(\alpha)$ $\mid (\exists h)$ $(su(\alpha, i, h)$ / $pto(h))$ $\geq$ \textit{interutil}\}\;
		compute \textit{Secondary}($\beta$) = $\{z \in $ \textit{Secondary}$(\alpha)$ $\mid (\exists h)$ $(lu(\alpha, i, h)$ / $pto(h))$ $\geq $ \textit{interutil}\}\;
		execute \textit{Search}($\beta$, $\beta$-$\mathcal{D}$, \textit{Primary}($\beta$), \textit{Secondary}($\beta$), \textit{interutil})\;
	}
\end{algorithm}

The \textit{Search} program (cf. Algorithm \ref{Algorithm:Search}) is called recursively. It requires a current prefix itemset $\alpha$, the projected database $\alpha-D$, the $Primary$ itemset of \textit{Primary}($\alpha$), the \textit{Secondary} itemset \textit{Secondary}($\alpha$) and the user-specified threshold $k$. The entire function traverses each item $i$ in the $Primary(\alpha)$. The algorithm then creates $\beta$ as a single-item extension of $\alpha$. The procedure scans the projected database $\alpha-D$ to compute the utility of $\beta$ and builds the projected database $\beta-D$ of $\beta$. Base on this projected database, it calculates the $ru(\beta)$. If $ru(\beta)$ is no less than the \textit{interutil}, the algorithm adds itemset $\beta$ into the set of top-$k$ itemsets. And it builds the \textit{NegativeItems}($\beta$) by the negative items which satisfied $(\exists h)(su(\beta, n, h)$ / $pto(h) >$ \textit{interutil}). And it calls the recursive function \textit{NegativeSearch} to add the negative item to the current itemset $\beta$. Then the procedure continues the scans according to each time period to calculate the $su$ and $lu$ of all items in \textit{Secondary}($\beta$). For all items $z \in $ \textit{Secondary}($\alpha$), once the algorithm finds a time period $h$ that makes $su(\beta, z, h)$/$pto(h))$ $\geq $ \textit{interutil} hold, the corresponding item $z$ must be added into \textit{Secondary}($\beta$). Similarly, once the algorithm finds a time period $h$ that makes $lu(\alpha, z, h)$/$pto(h)) \geq $ \textit{interutil} hold, the corresponding item $z$ must also be added into \textit{Secondary}($\beta$). Finally, this procedure calls itself recursively.

TOIT uses the \textit{NegativeSearch} function (cf. Algorithm \ref{Algorithm:NSearch}) to extend the current prefix itemset using negative items in the database. This function requires the current prefix itemset $\alpha$, the projected database $\alpha-D$, the \textit{NegativeItems} itemset of $\alpha$, and the user-specified threshold $k$. It iterates over each item $i$ in \textit{NegativeItems}($\alpha$). Then it builds the single-item expansion $\gamma$ = $\alpha \cup \{z\}$ and then calculates the $ru$ of $\gamma$. If $ru(\gamma)$ $\geq$ \textit{interutil} holds, it adds itemset $\gamma$ into the set of top-$k$ itemsets. The procedure updates the value of $su$ for each item which is in \textit{NegativeItems}($\alpha$). If there is a time period $h$ that meets the formula $q(z)$ $< 0$ $\wedge$ $su(\gamma, z, h)$/$pto(h)) \geq$ \textit{interutil}, this negative item $z$ would be added into \textit{NegativeItems}($\gamma$). In the end, this procedure calls itself recursively.

% NSearch Algorithm
\begin{algorithm}[h]
	\caption{The NegativeSearch procedure}
	\label{Algorithm:NSearch}
	\KwIn{
		$\alpha$: a prefix itemset, $\alpha$-$\mathcal{D}$: the projected database based on $\alpha$, \textit{NegativeItems}($\alpha$): the promising negative items of $\alpha$, $k$: a user-specified threshold amount of pattern.
	}
	\KwOut{a set of itemsets which are supersets of $\alpha$ with on-shelf high utility.
	}
	
	\For{\rm each item $i \in $ \textit{NegativeItems}($\alpha$)}{
		$\gamma$ = $\alpha \cup\{i\}$\;
		calculate $ru(\gamma)$\;
		\If{$ru(\gamma) \geq$ \textit{interutil}}{
			add ItemsetToTopK($\gamma$)\;
		}
		
		\For{\rm each time period $h$}{
			\For{\rm each item $z \in $ \textit{NegativeItems}($\alpha$) in $h$}{
				update the $su(\gamma, z, h)$ \;
			}
		}
		
		compute \textit{NegativeItems}($\gamma$) = $\{n \mid q(n) < 0 \wedge (\exists h) (su(\gamma, n, h)$ / $pto(h)$ $\geq$ \textit{interutil})\}\;
		call \textit{NegativeSearch}($\gamma$, $ \gamma$-$\mathcal{D}$, \textit{NegativeItems}($\beta$), $k$)\;
	}
\end{algorithm}

The $RIU$ procedure (cf. Algorithm \ref{Algorithm:RIU}) rapidly raises the \textit{interutil} at the start. This procedure requires the original database $\mathcal{D}$ and the user-specified threshold $k$. First, it calculates the relative utility $ru$ of each single-item itemset. Then, if the amount of $ru$ which is greater than or equal to $0$ is no less than $k$, it sets the \textit{interutil} to the value of $k$-th high $ru$. Otherwise, if the amount of $ru$ which is greater than or equal to 0 is less than $k$, it sets the \textit{interutil} to $0$.

% RIU Algorithm
\begin{algorithm}[h]
	\caption{The RIU procedure}
	\label{Algorithm:RIU}
	\KwIn{$\mathcal{D}$: database, $k$: a user-specified threshold amount of pattern.}
	\KwOut{\textit{interutil}: the value may equal to the relative utility of $k$-th itemset if the amount of patterns is no less than $k$ at the end.}
	
	calculate $ru(i)$ of each item $i \in D$\;
	
	\eIf{\rm the amount of $ru$ (which $\geq$ 0) is no less than $k$}
	{set \textit{interutil} equal to the $k$-th high $ru$\;}
	{set \textit{interutil} = 0\;}
\end{algorithm}

\section{Experiments}
\label{sec:experiments}

To accurately evaluate the performance of the TOIT algorithm, we have conducted extensive experiments on various datasets and compared them with the state-of-the-art KOSHU algorithm \cite{dam2017efficient}. It should be pointed out that the source code of KOSHU will miss some interesting patterns in the final results. Thus, we debugged the source code of the KOSHU algorithm to ensure its accuracy before testing. All experiments are carried out on a computer with a Windows 10 operating system, a 64-bit Inter Core i7, 3.0 GHz, and 16 GB of memory. We implemented TOIT and KOSHU in the Java language. For reproducibility, we make all source code available on GitHub {\color{blue} https://github.com/DSI-Lab1/TOIT}. In our tests, the maximum memory of the JVM was set to 10 GB. In addition, if the execution time of an experiment is more than 10,000 seconds, we assume it cannot get results within a reasonable time and do not draw it in line charts.

All tested datasets are downloaded from the SPMF open source library\footnote{\url{http://www.philippe-fournier-viger.com/spmf/index.php}}. Table \ref{talbe:datasetcharacter} lists the characteristics of these datasets. The accidents dataset has the largest number of transactions, and the average length of each transaction is also very long. The chess is a dense dataset with long transactions but few items, and its average length is the longest one in the experiment. The last dense dataset is mushroom, and its characteristic is moderation. The retail dataset has sparse features. It has the largest number of distinct items, but with the shortest average length of each transaction. Furthermore, in order to know the impact of different quantities of time periods on the mining performance, we set 4, 8, and 12 time periods for each dataset. Except for the time period, the components of transactions were thus fixed.

\begin{table*}[h]
	\caption{The characteristics of datasets}
	\label{talbe:datasetcharacter}
	\centering
	\begin{tabular}{|c|c|c|c|c|c|}
		\hline
		\textbf{Dataset}	&	Transactions	&	Distinct items	&	AvgLen	&	MaxLen	& Period\\	\hline \hline
		\textbf{accident}	&	340,183	&	468	&	33.8	&	51	&	4 $\sim$ 12 \\	\hline
		\textbf{chess}		&	3,196	&	75	&	37	&	37	&	4 $\sim$ 12 \\	\hline
		\textbf{mushroom}	&	8,124	&	119	&	23	&	23	&	4 $\sim$ 12 \\	\hline
		\textbf{retail}		&	88,162	&	16,470	&	10.3	&	76	&	4 $\sim$ 12 \\	\hline
	\end{tabular}
\end{table*}

\subsection{Influence of the Top-$k$ Threshold on Time Execution}

\begin{figure}   	
	% accidents
	\begin{minipage}[t]{0.3\textwidth}	
		\centering   	
		\includegraphics[scale=0.35, trim=100 0 0 0]{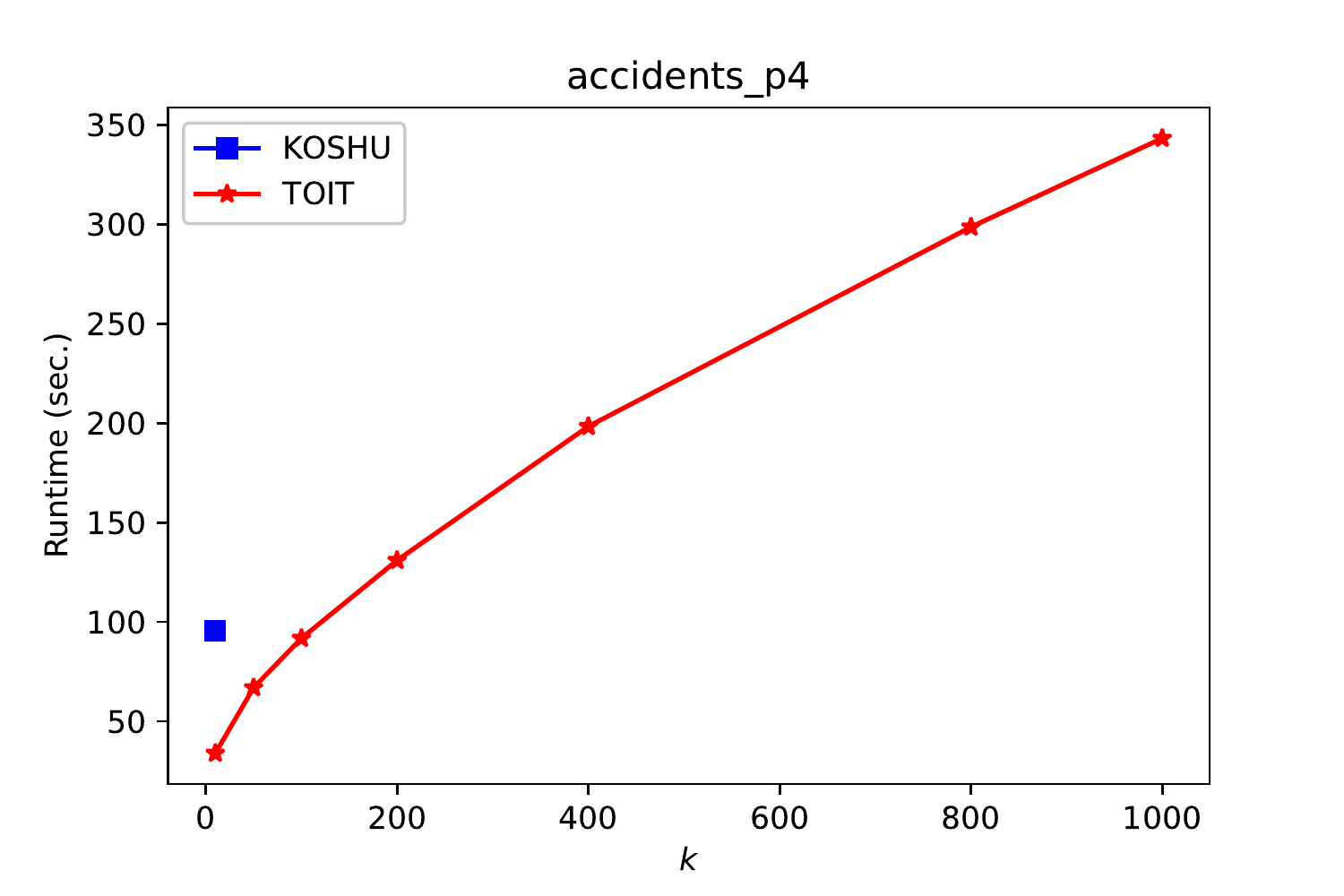}
	\end{minipage}
	\begin{minipage}[t]{0.3\textwidth}   	
		\centering   	
		\includegraphics[scale=0.35, trim=50 0 0 0]{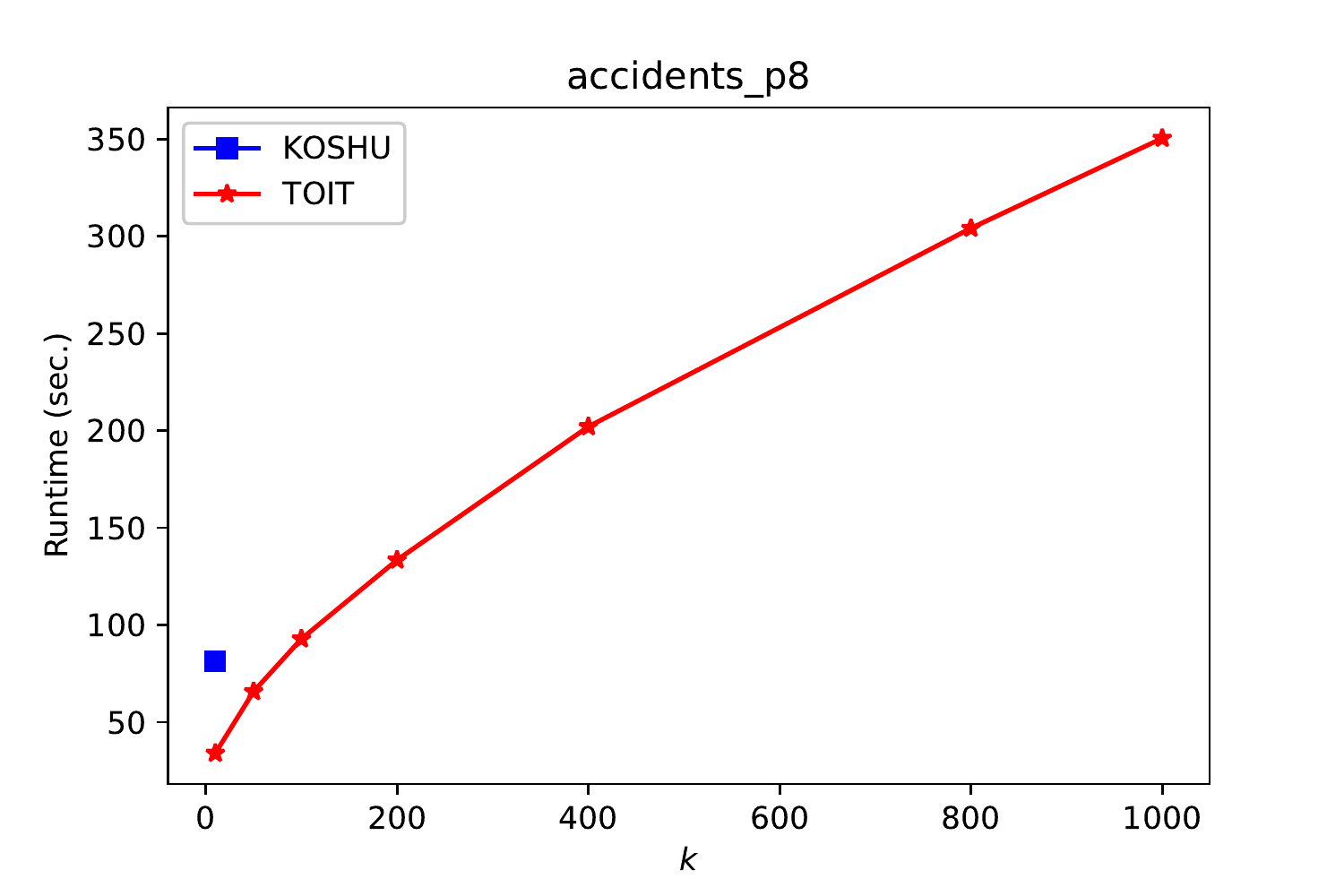}   	
	\end{minipage}
	\begin{minipage}[t]{0.3\textwidth}   	
		\centering   	
		\includegraphics[scale=0.35]{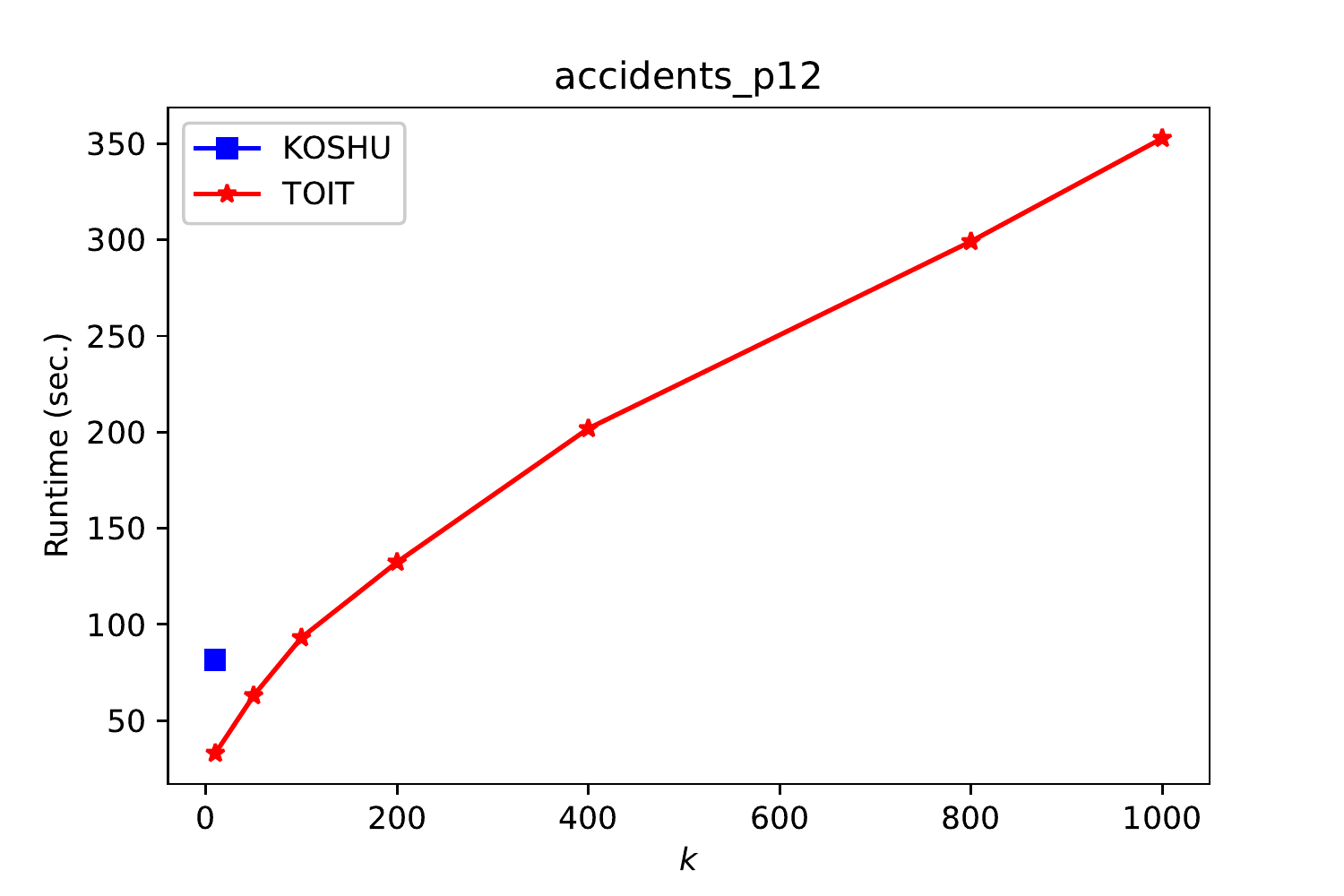}
	\end{minipage}
	
	% chess
	\begin{minipage}[t]{0.3\textwidth}	
		\centering   	
		\includegraphics[scale=0.35, trim=100 0 0 0]{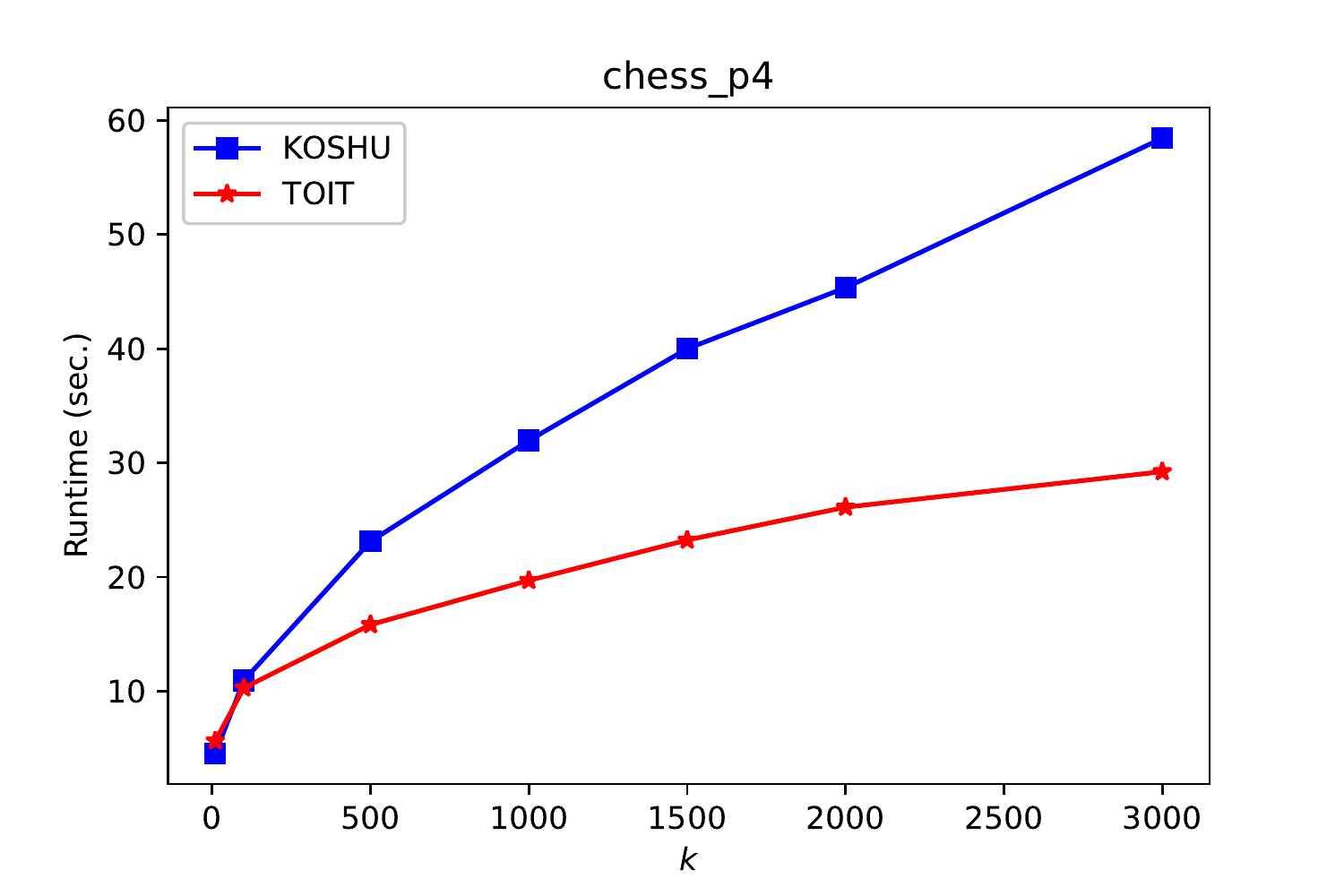}
	\end{minipage}
	\begin{minipage}[t]{0.3\textwidth}   	
		\centering   	
		\includegraphics[scale=0.35, trim=50 0 0 0]{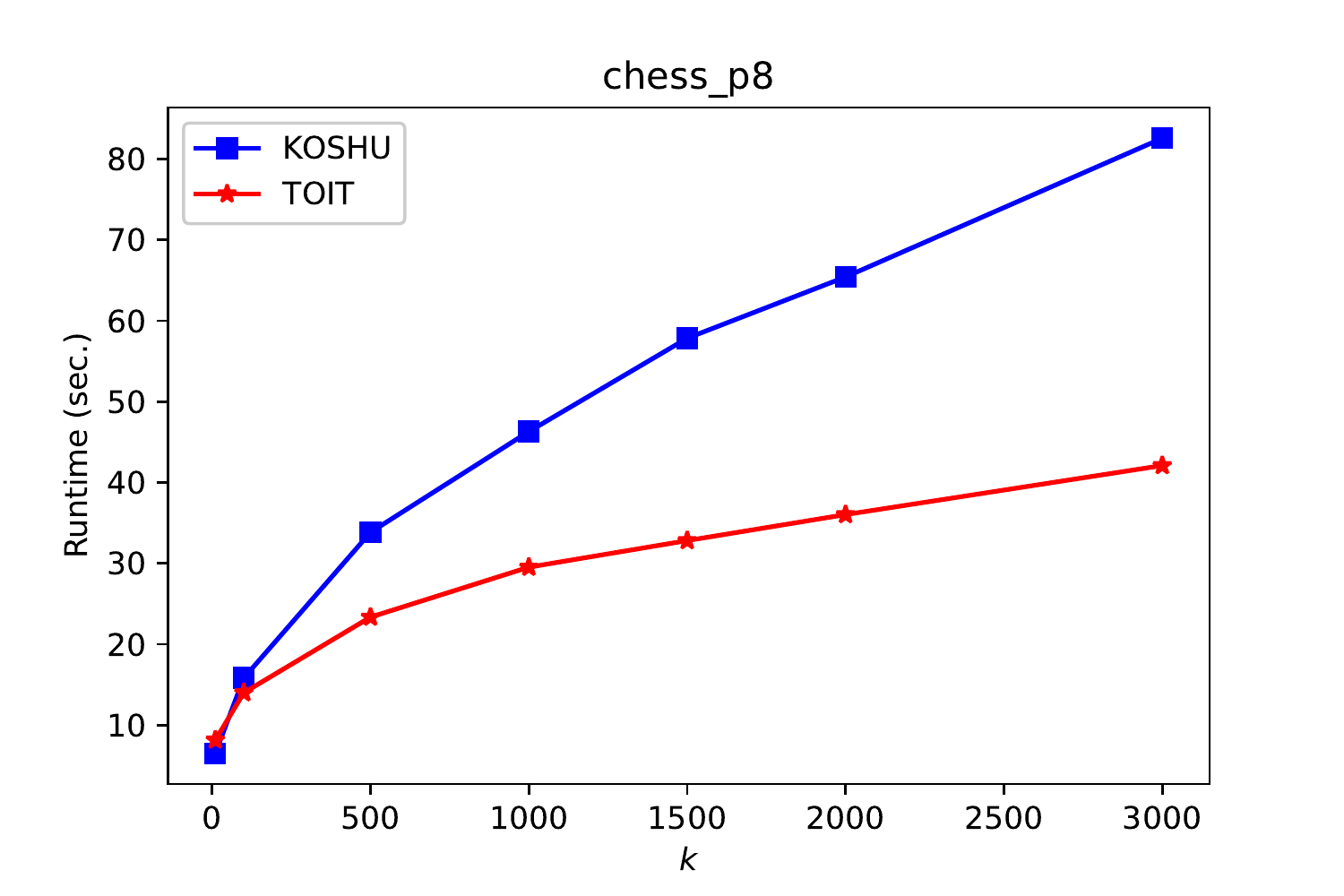}   	
	\end{minipage}
	\begin{minipage}[t]{0.3\textwidth}   	
		\centering   	
		\includegraphics[scale=0.35]{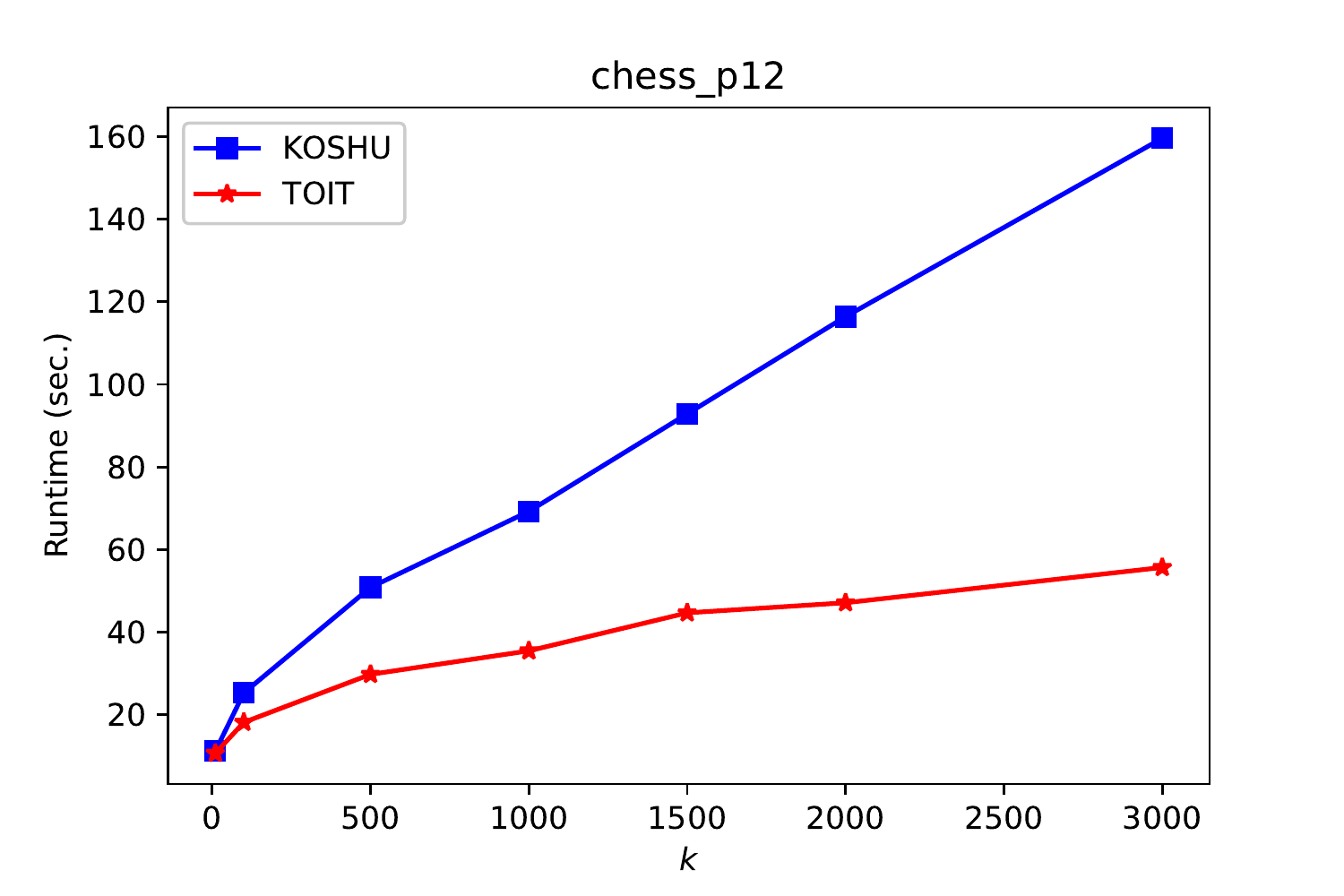}
	\end{minipage}
	
	% mushroom
	\begin{minipage}[t]{0.3\textwidth}	
		\centering   	
		\includegraphics[scale=0.35, trim=100 0 0 0]{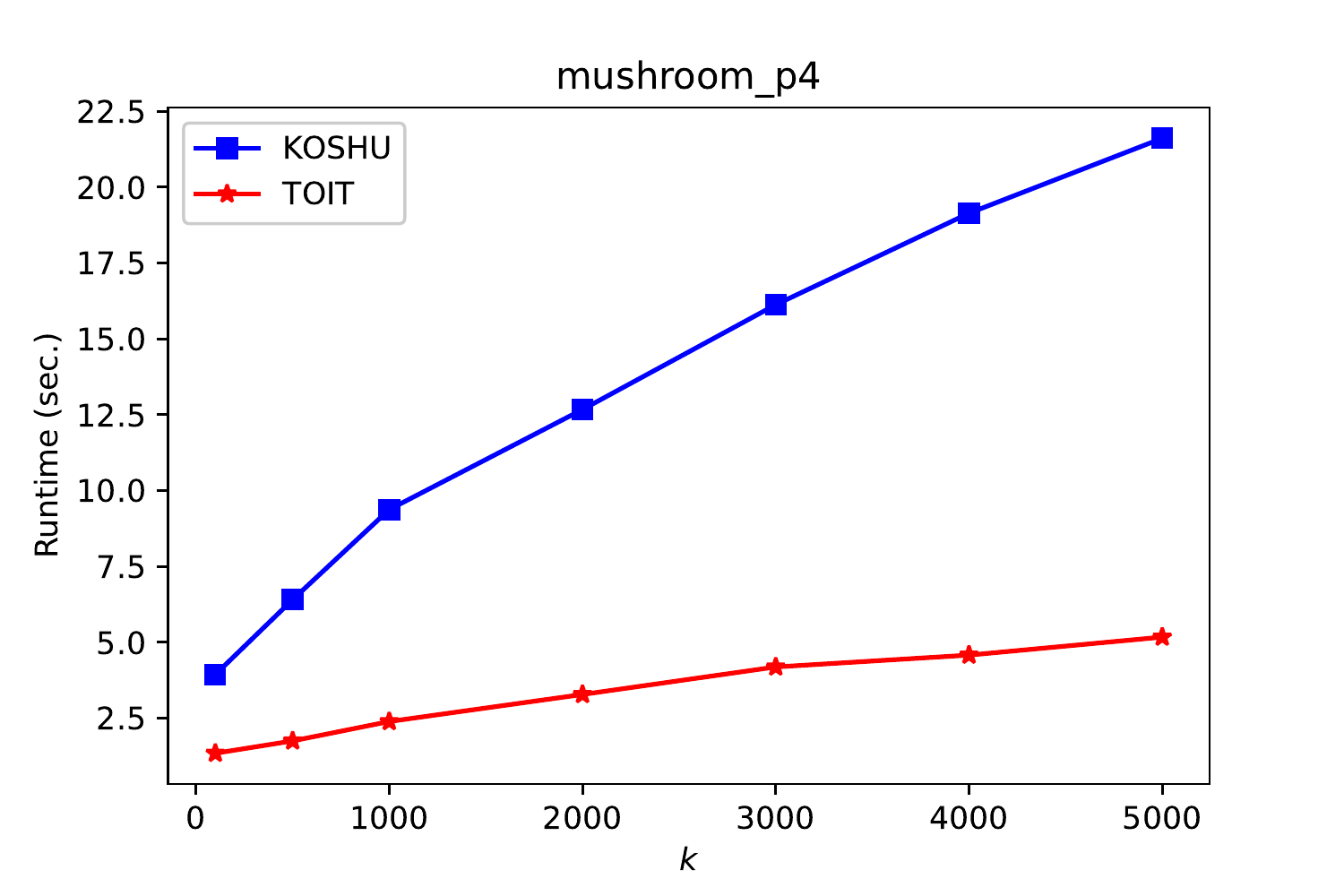}

	\end{minipage}
	\begin{minipage}[t]{0.3\textwidth}   	
		\centering   	
		\includegraphics[scale=0.35, trim=50 0 0 0]{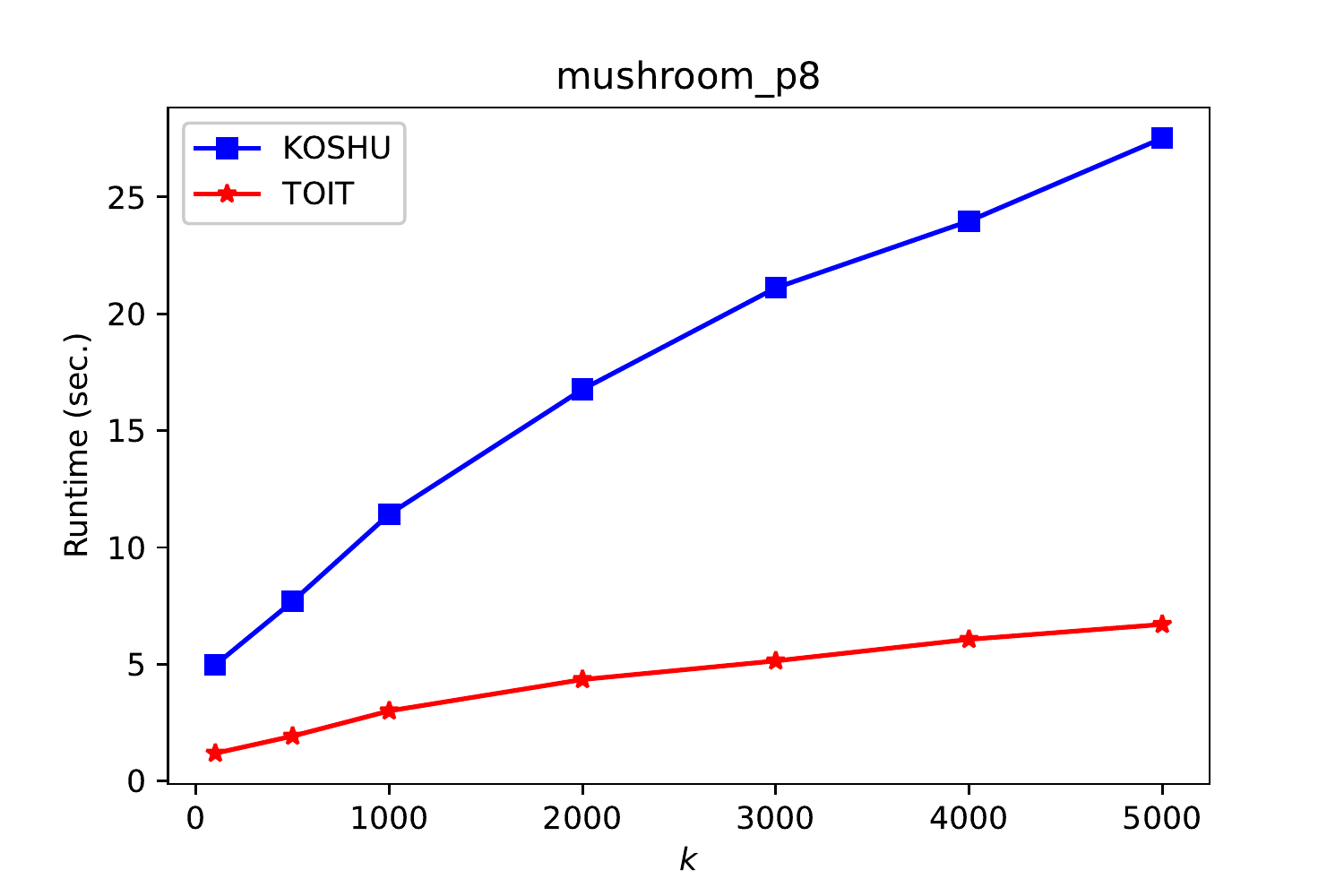}   	
	\end{minipage}
	\begin{minipage}[t]{0.3\textwidth}   	
		\centering   	
		\includegraphics[scale=0.35]{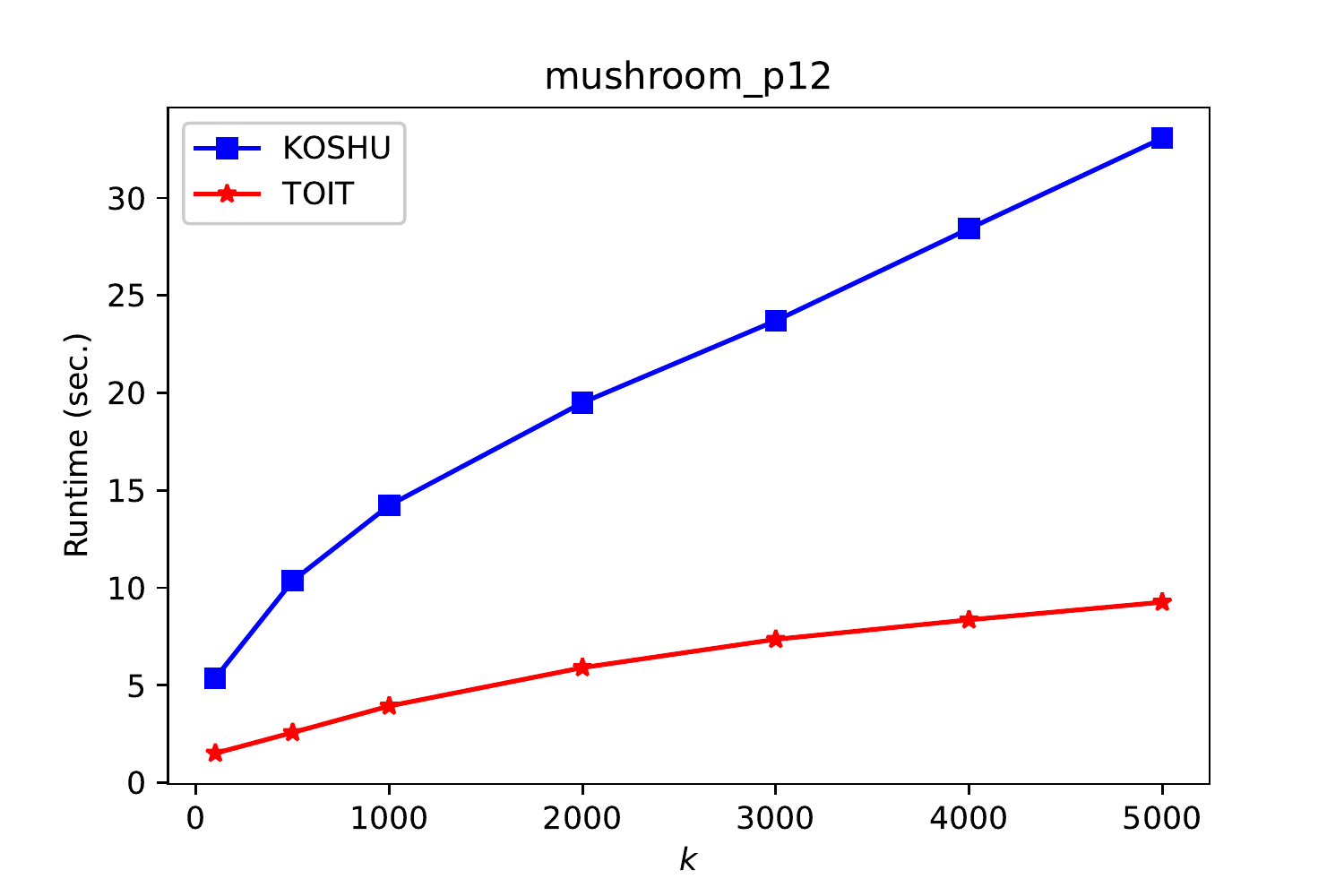}
	\end{minipage}
	
	% retail
	\begin{minipage}[t]{0.3\textwidth}	
		\centering   	
		\includegraphics[scale=0.35, trim=100 0 0 0]{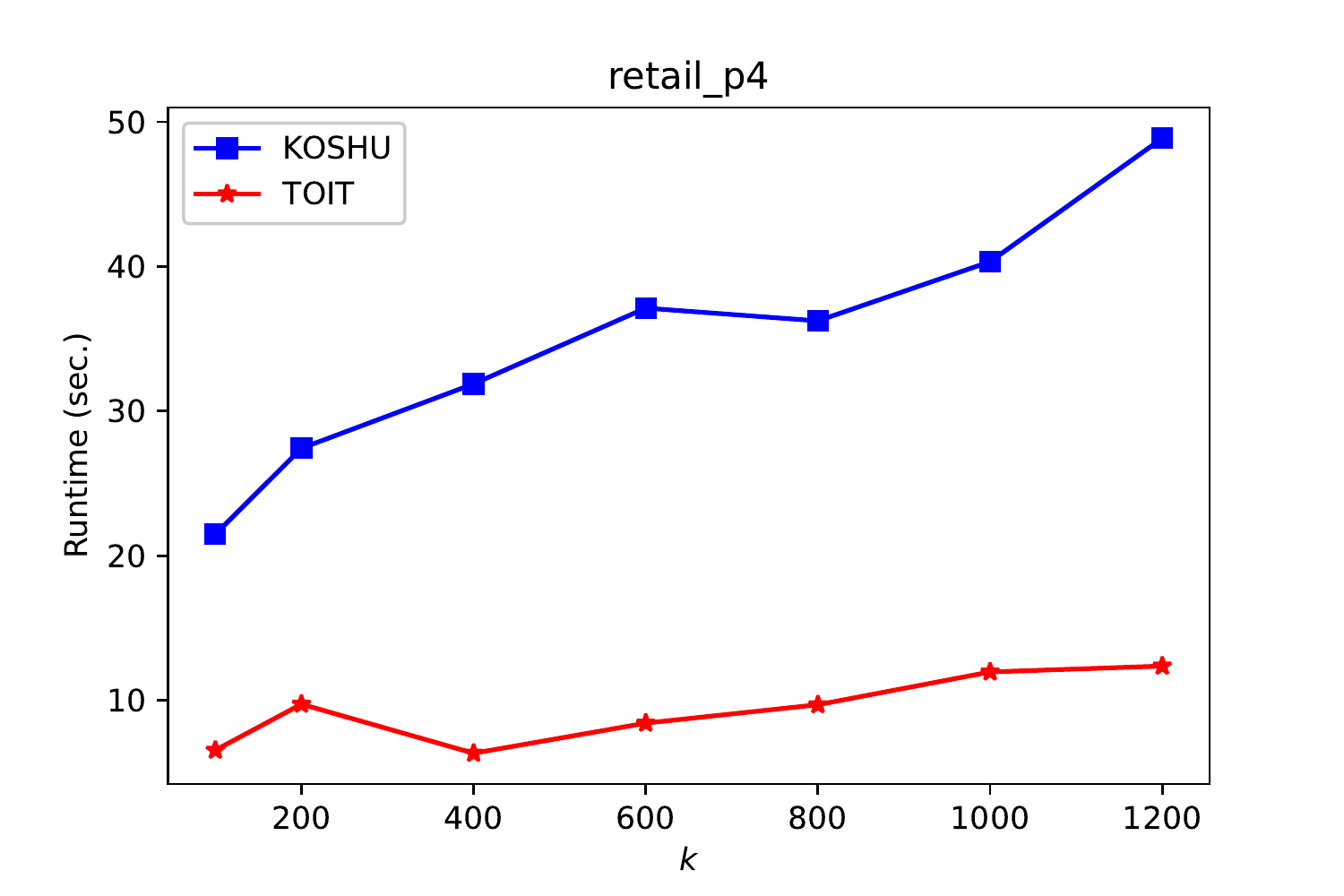}
	\end{minipage}
	\begin{minipage}[t]{0.3\textwidth}   	
		\centering   	
		\includegraphics[scale=0.35, trim=50 0 0 0]{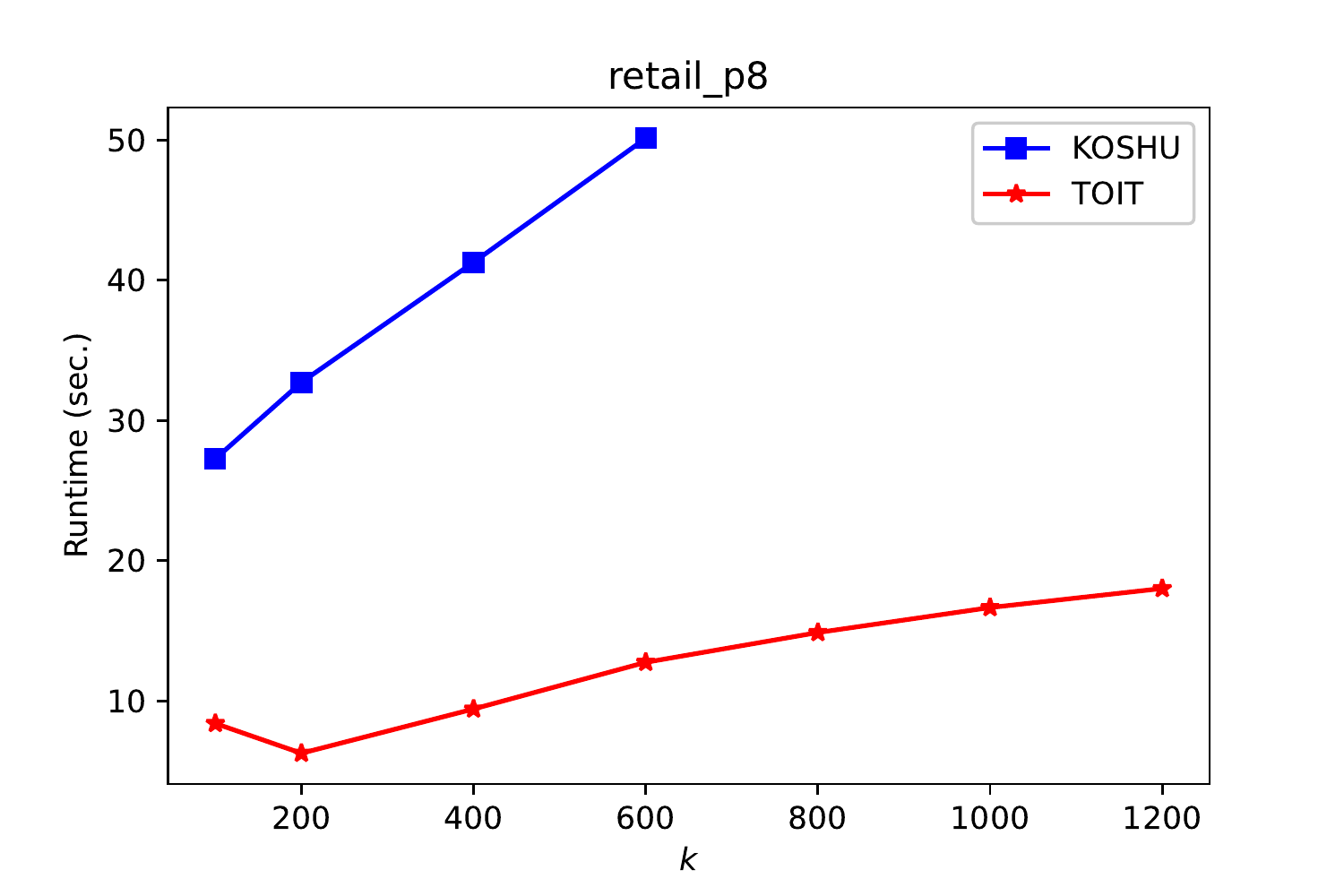}   	
	\end{minipage}
	\begin{minipage}[t]{0.3\textwidth}   	
		\centering   	
		\includegraphics[scale=0.35]{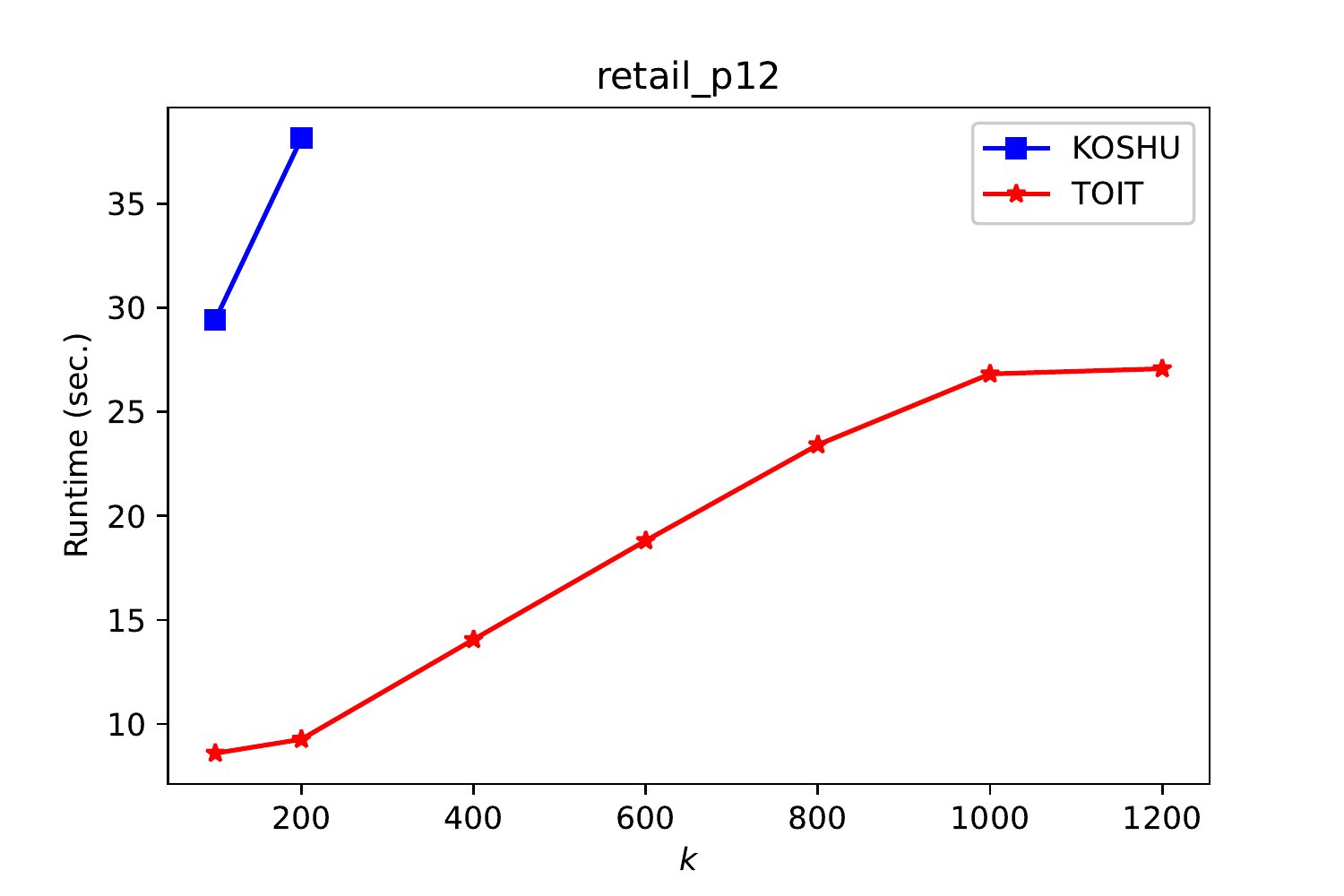}
	\end{minipage}
	
	\caption{Runtime consumption on different datasets with various $k$}
	\label{fig:runtime}	
\end{figure}

We began by testing execution time with various $k$ values. As illustrated in Fig. \ref{fig:runtime}, the time consumption of KOSHU and TOIT both increases as $k$ increases, and the runtime gap between the two algorithms widens. Generally, the runtime of TOIT is around 2x to 6x faster than that of KOSHU in most datasets. However, due to the dense nature of the accident dataset, KOSHU can only obtain results when $k$ is close to 0 in 4, 8, and 12 time periods, and TOIT's time consumption also increases rapidly. Considering the sparse dataset (i.e., retail), when the time period is 12 and $k$ is set to 200, TOIT takes nearly 10 seconds while KOSHU needs over 37 seconds. Unfortunately, while $k$ is higher than 200, KOSHU always takes too long to run with 12 time periods. Due to the maximum $k$ value of 5,000, the volatility of the range of time consumption is very small except in the chess dataset. On the chess dataset, when time period is 4, the range of KOSHU's runtime is around 2 to 58 seconds, but it is nearly 10 to 159 seconds when time period is 12. What's more, compared with KOSHU, the proposed algorithm always needs 2 to 30 seconds, which is more stable. All in all, the execution time experiments demonstrate that the TOIT is saving more time than KOSHU needs in various datasets. In particular, TOIT has a good performance on runtime in sparse datasets.

\begin{figure}   	
	% accidents
	\begin{minipage}[t]{0.3\textwidth}	
	\centering   	
	\includegraphics[scale=0.35, trim=100 0 0 0]{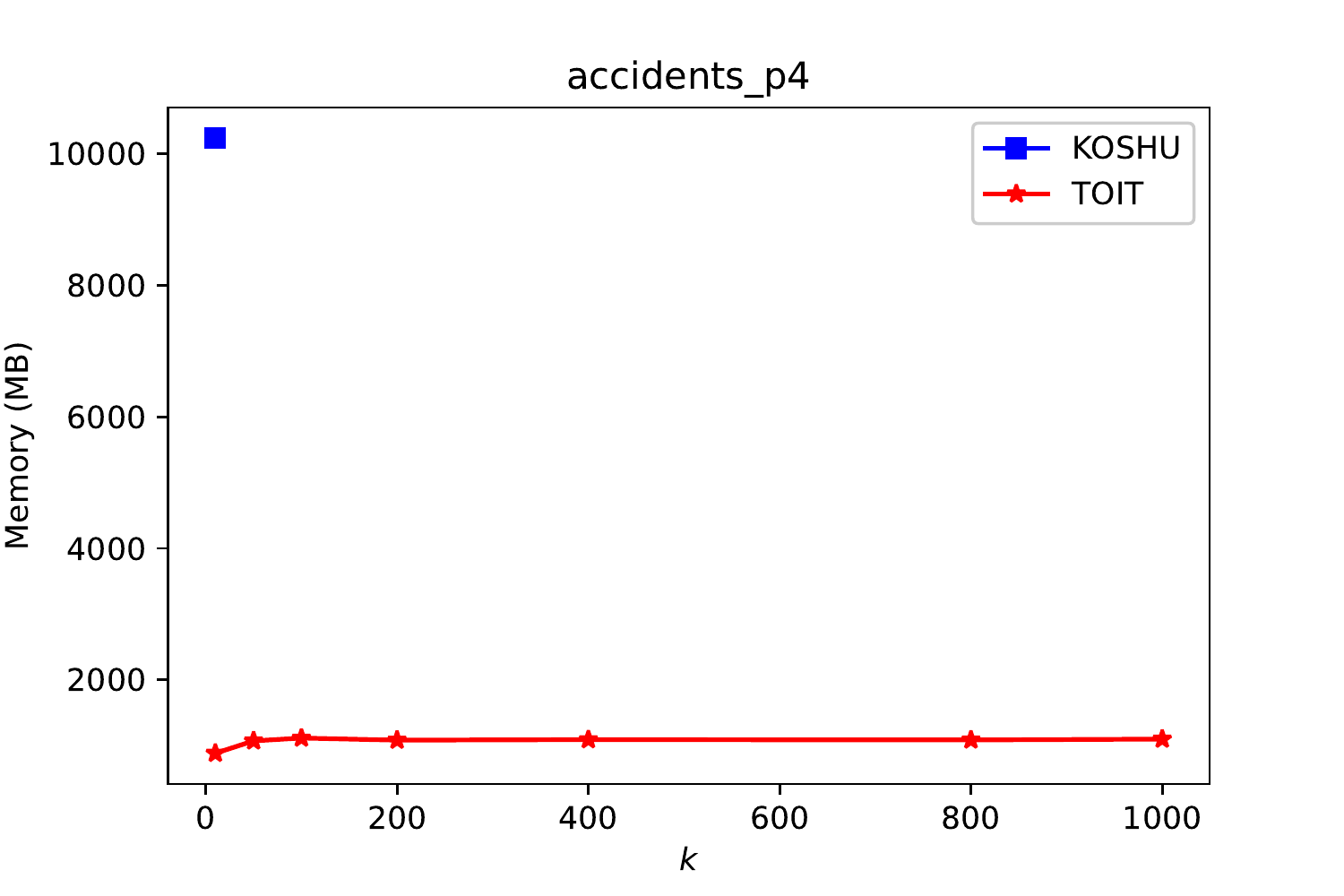}
	\end{minipage}
	\begin{minipage}[t]{0.3\textwidth}   	
		\centering   	
		\includegraphics[scale=0.35, trim=50 0 0 0]{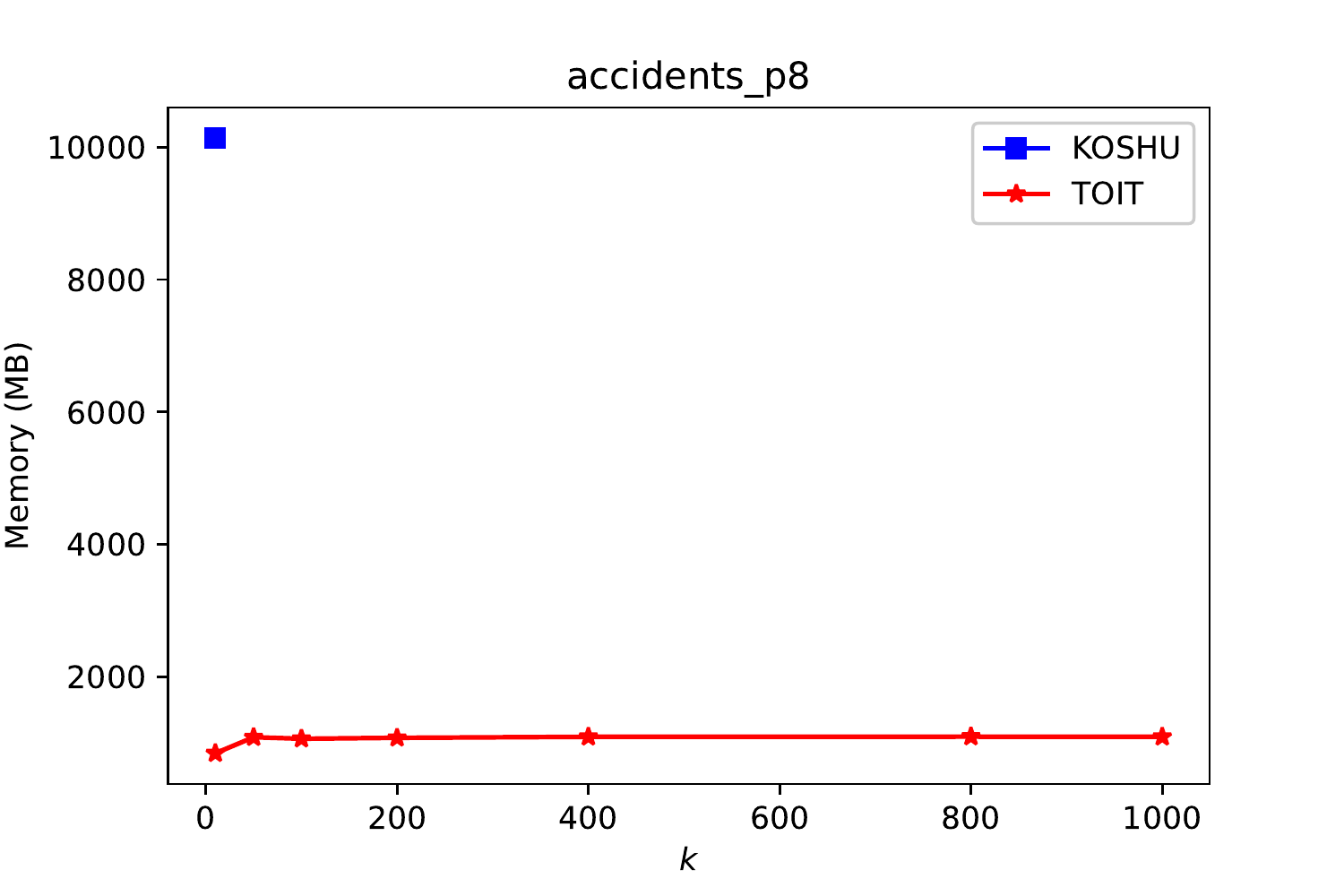}   
	\end{minipage}
	\begin{minipage}[t]{0.3\textwidth}   	
		\centering   	
		\includegraphics[scale=0.35]{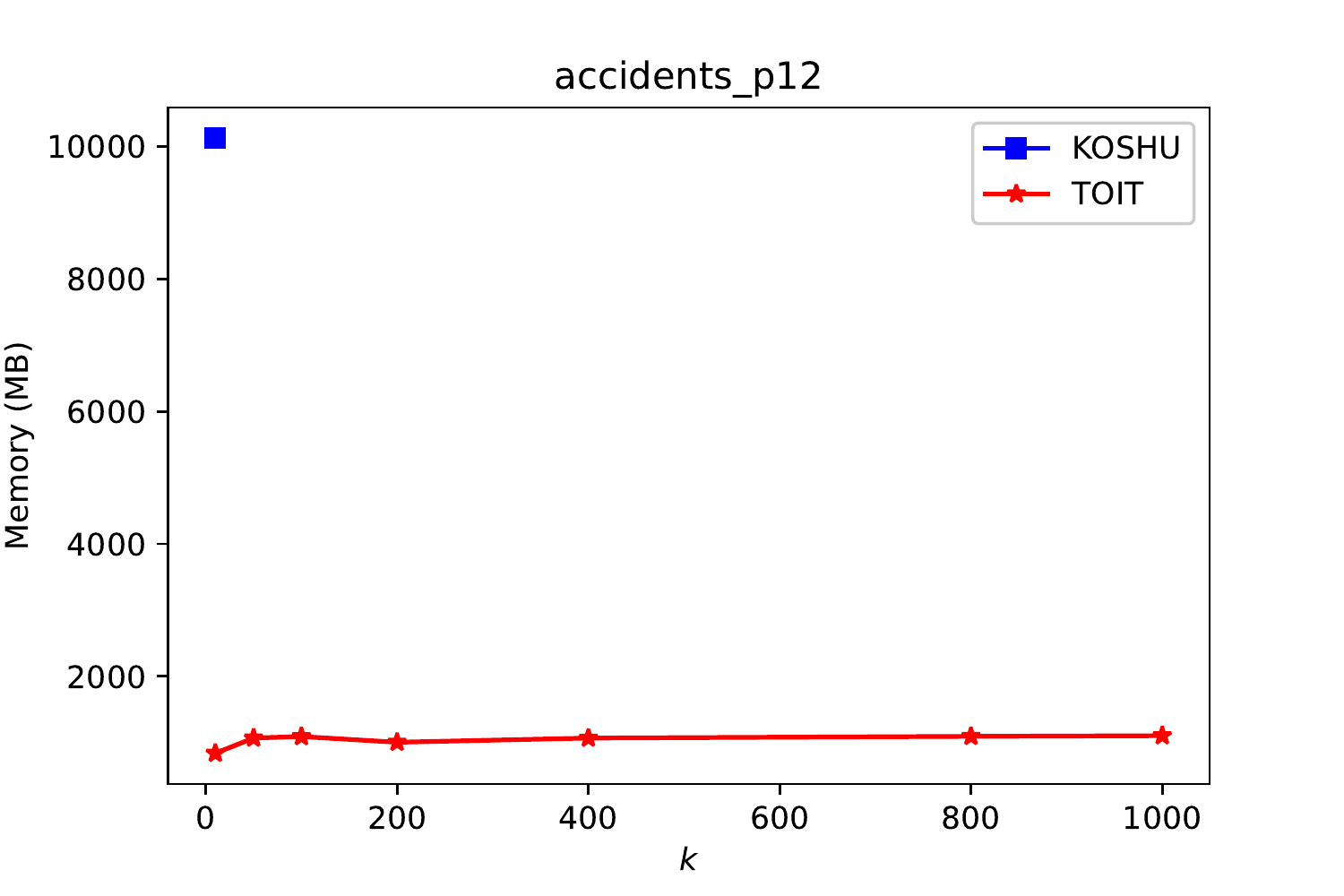}
	\end{minipage}
	
	% chess
	\begin{minipage}[t]{0.3\textwidth}	
		\centering   	
		\includegraphics[scale=0.35, trim=100 0 0 0]{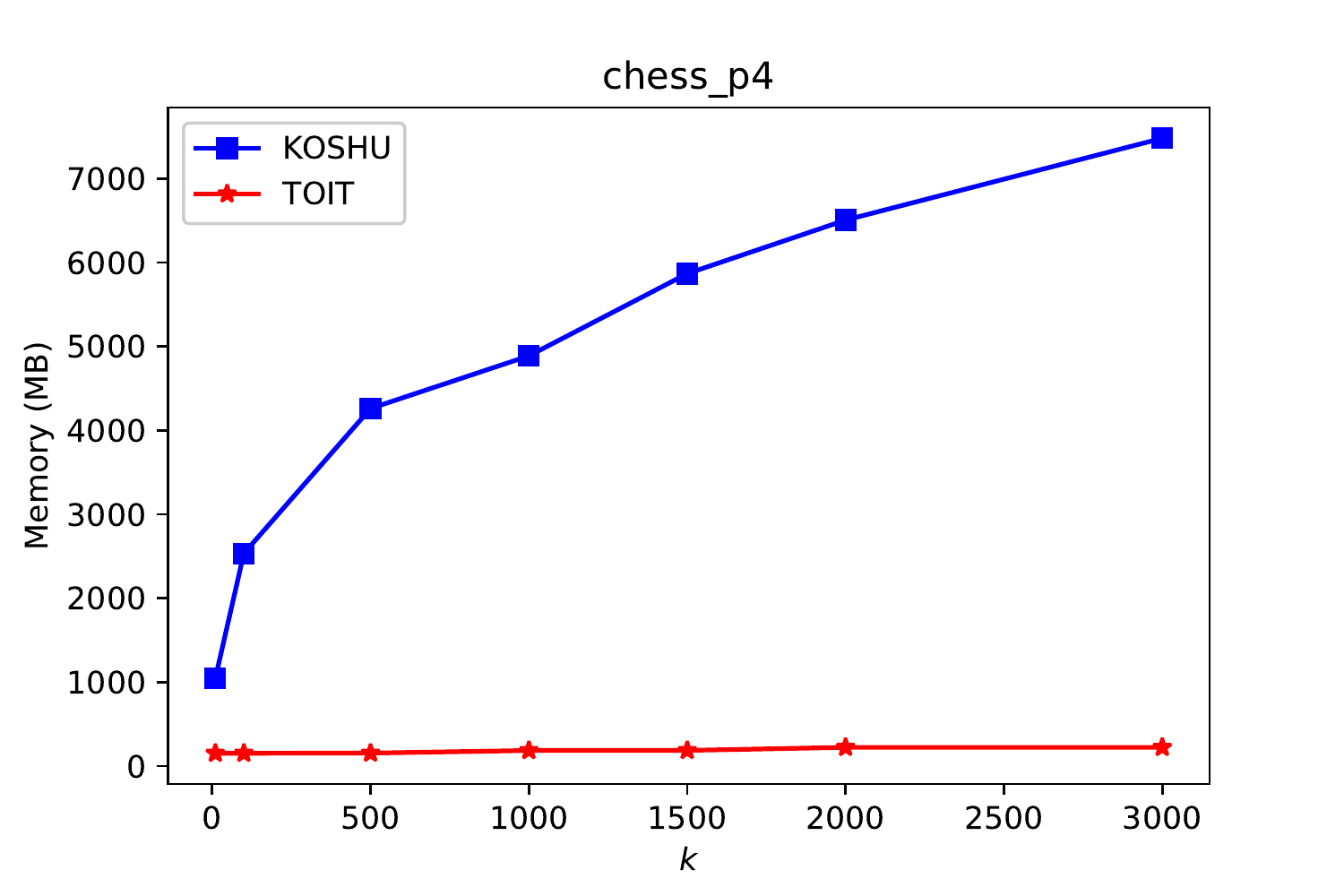}
	\end{minipage}
	\begin{minipage}[t]{0.3\textwidth}   	
		\centering   	
		\includegraphics[scale=0.35, trim=50 0 0 0]{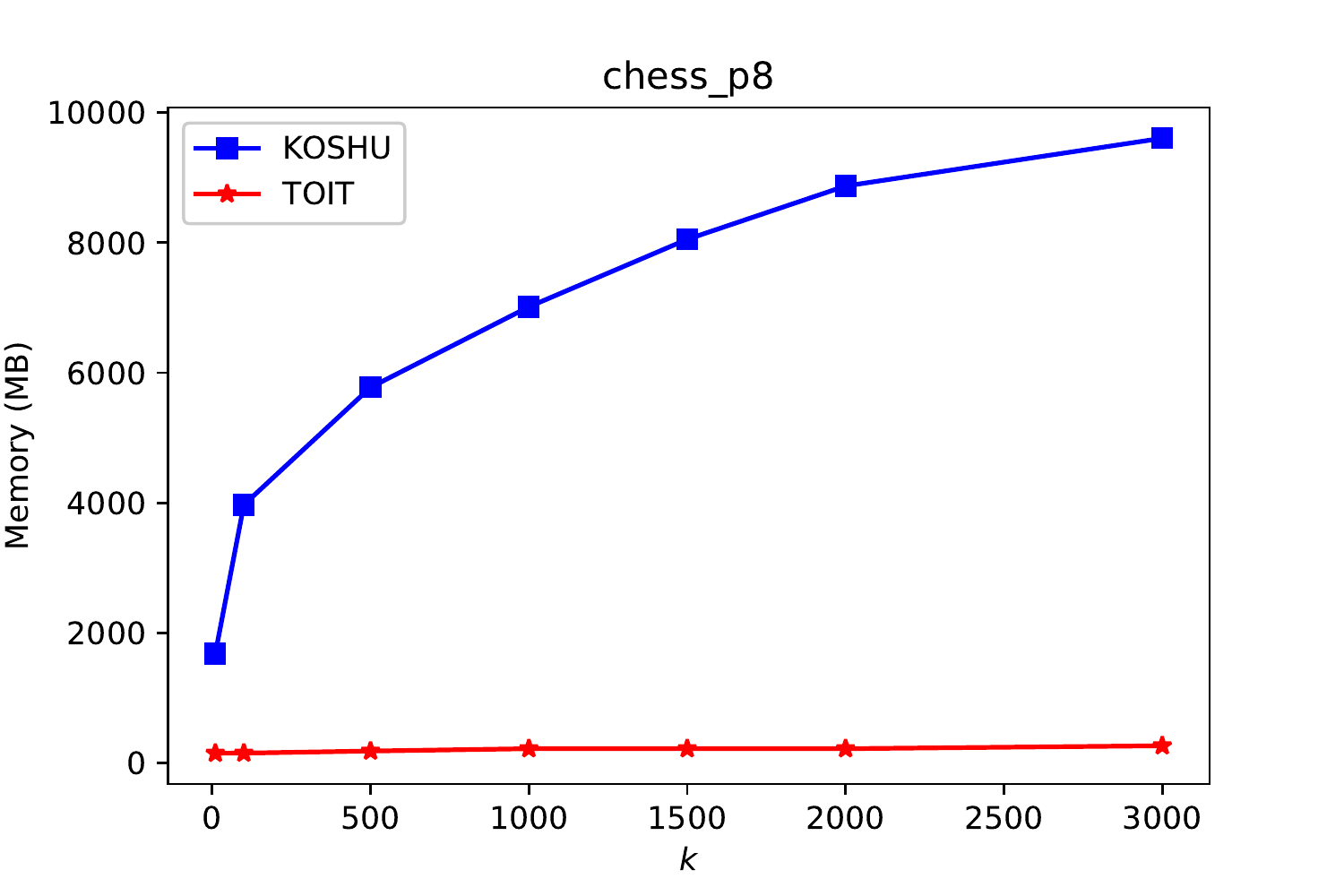}   	
	\end{minipage}
	\begin{minipage}[t]{0.3\textwidth}   	
		\centering   	
		\includegraphics[scale=0.35]{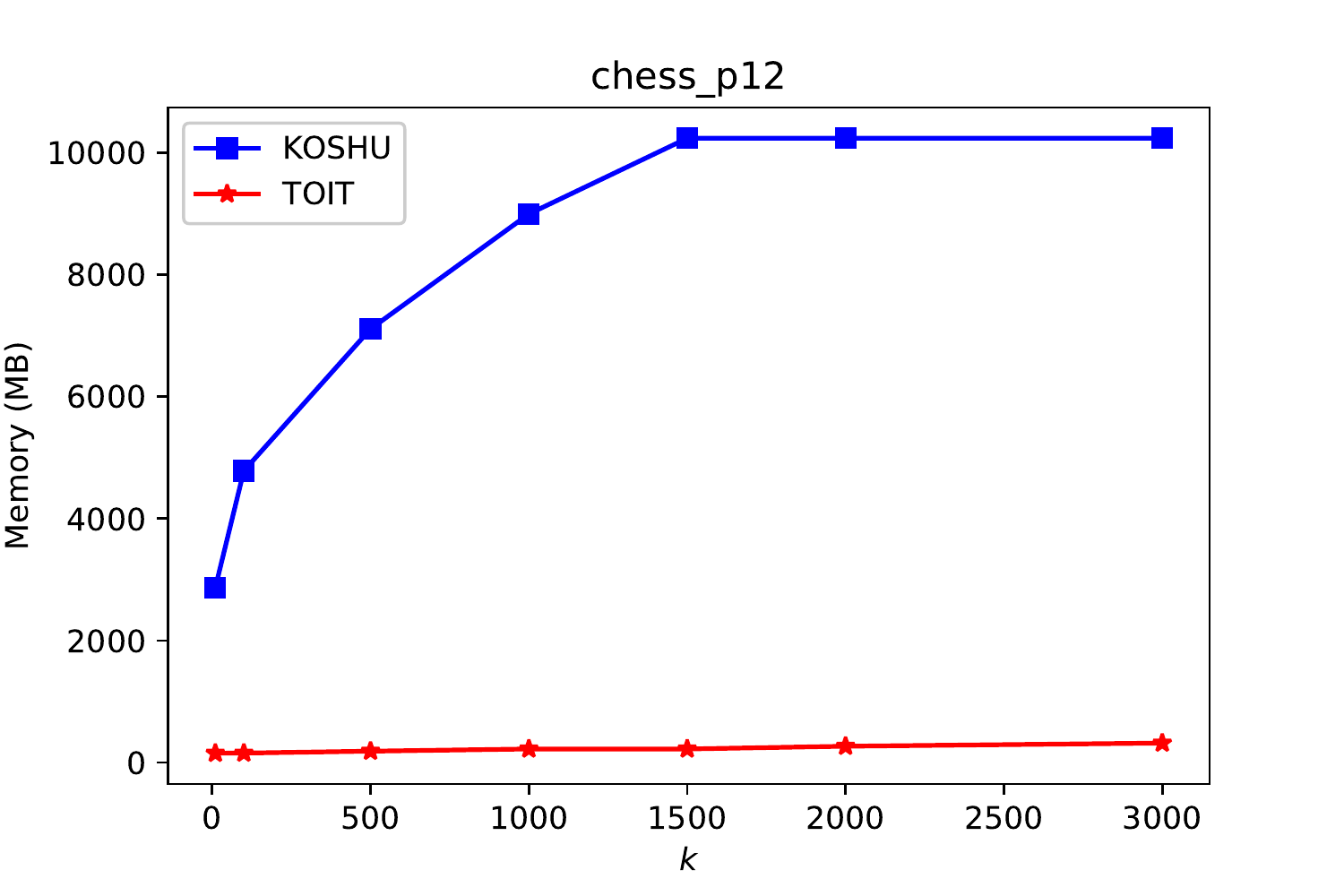}
	\end{minipage}
	
	% mushroom
	\begin{minipage}[t]{0.3\textwidth}	
		\centering   	
		\includegraphics[scale=0.35, trim=100 0 0 0]{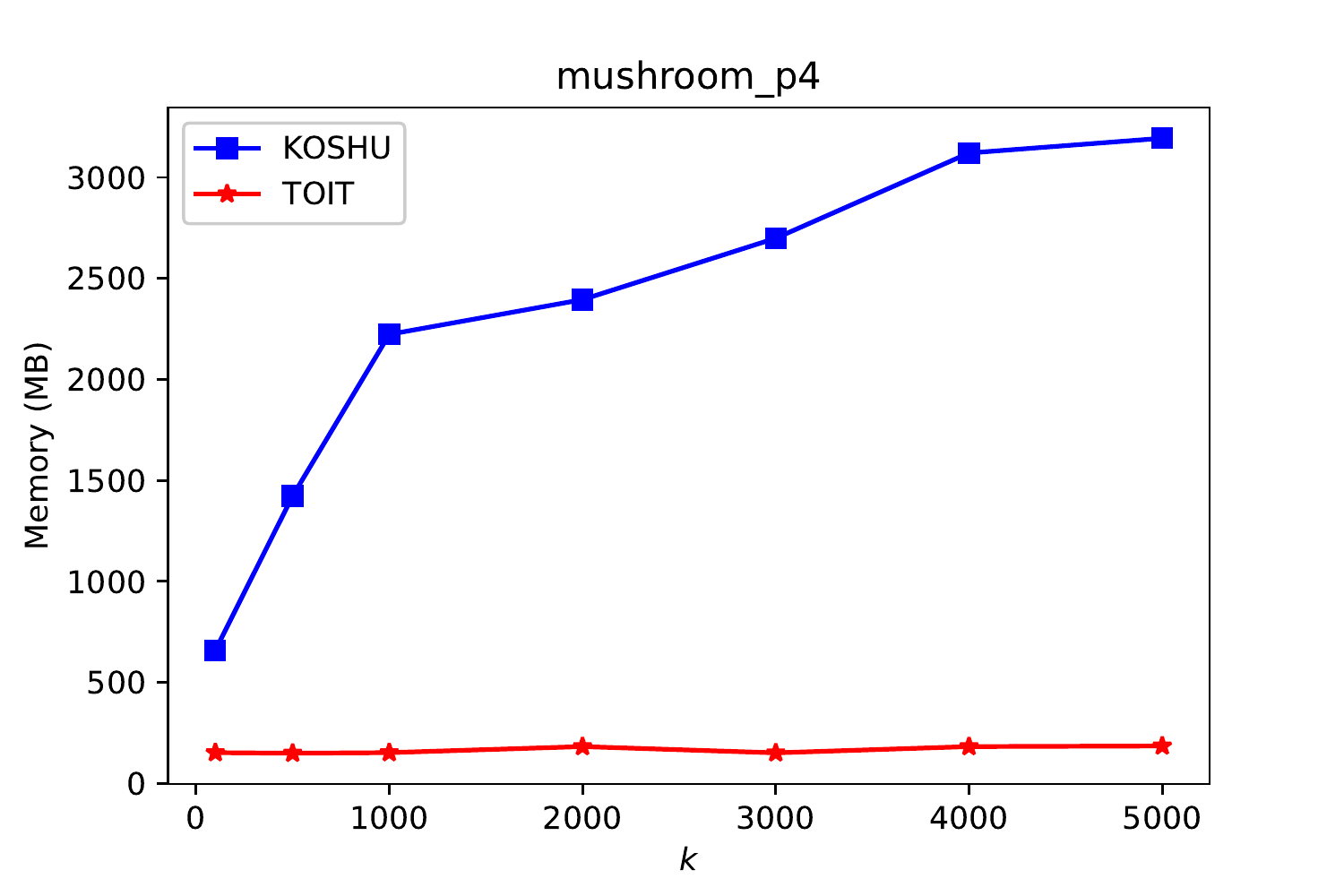}
	\end{minipage}
	\begin{minipage}[t]{0.3\textwidth}   	
		\centering   	
		\includegraphics[scale=0.35, trim=50 0 0 0]{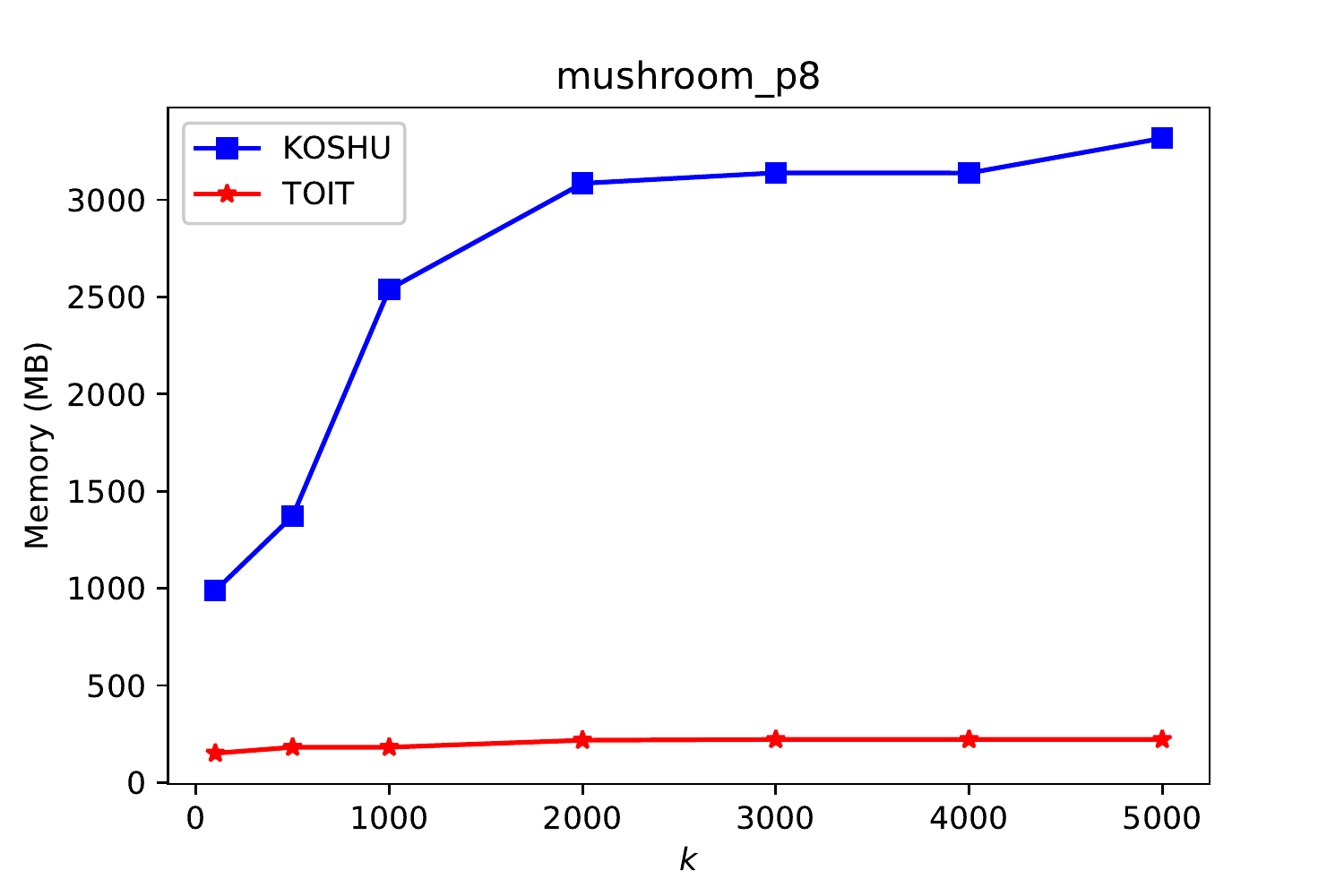}   
	\end{minipage}
	\begin{minipage}[t]{0.3\textwidth}   	
		\centering   	
		\includegraphics[scale=0.35]{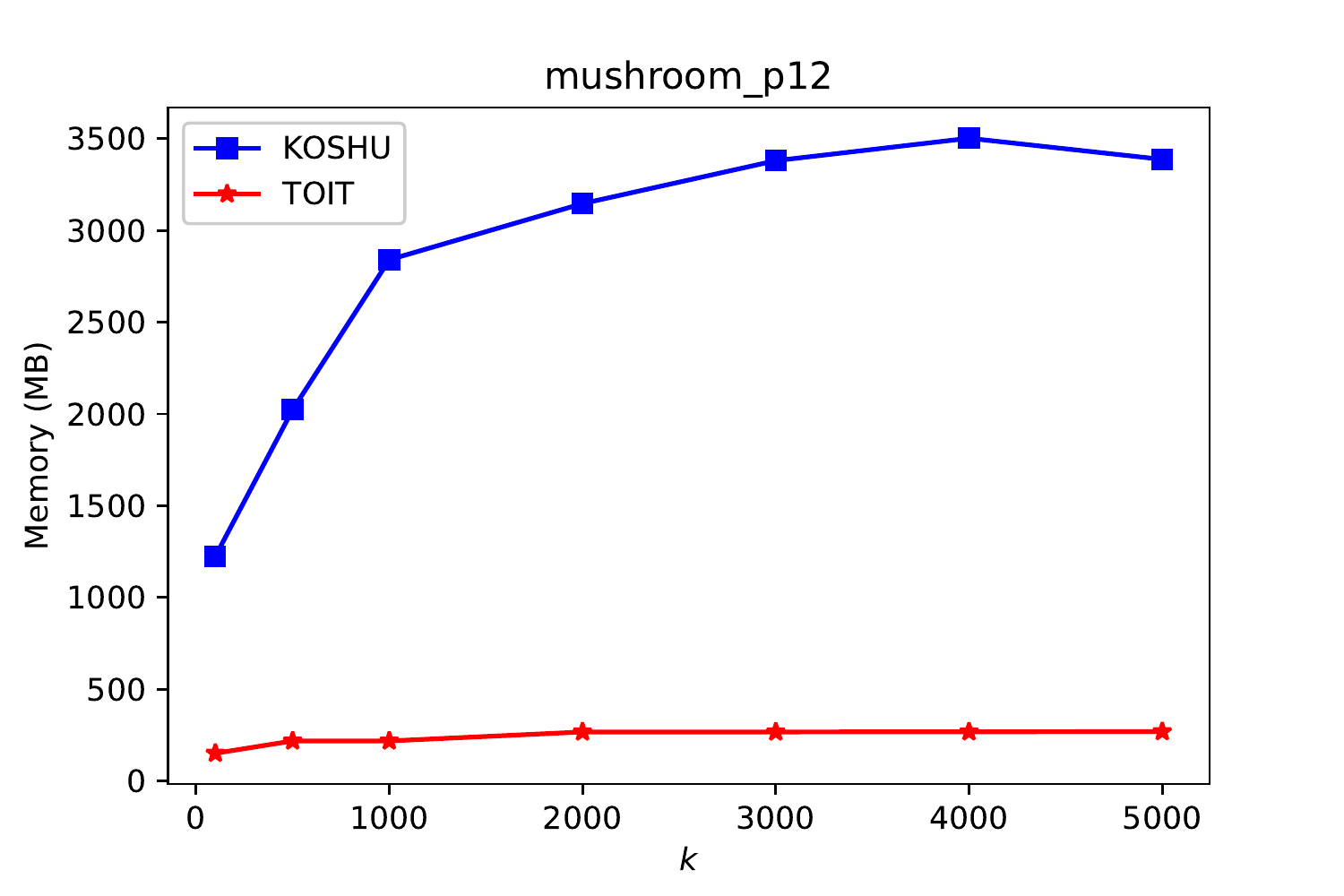}
	\end{minipage}
	
	% retail
	\begin{minipage}[t]{0.3\textwidth}	
		\centering   	
		\includegraphics[scale=0.35, trim=100 0 0 0]{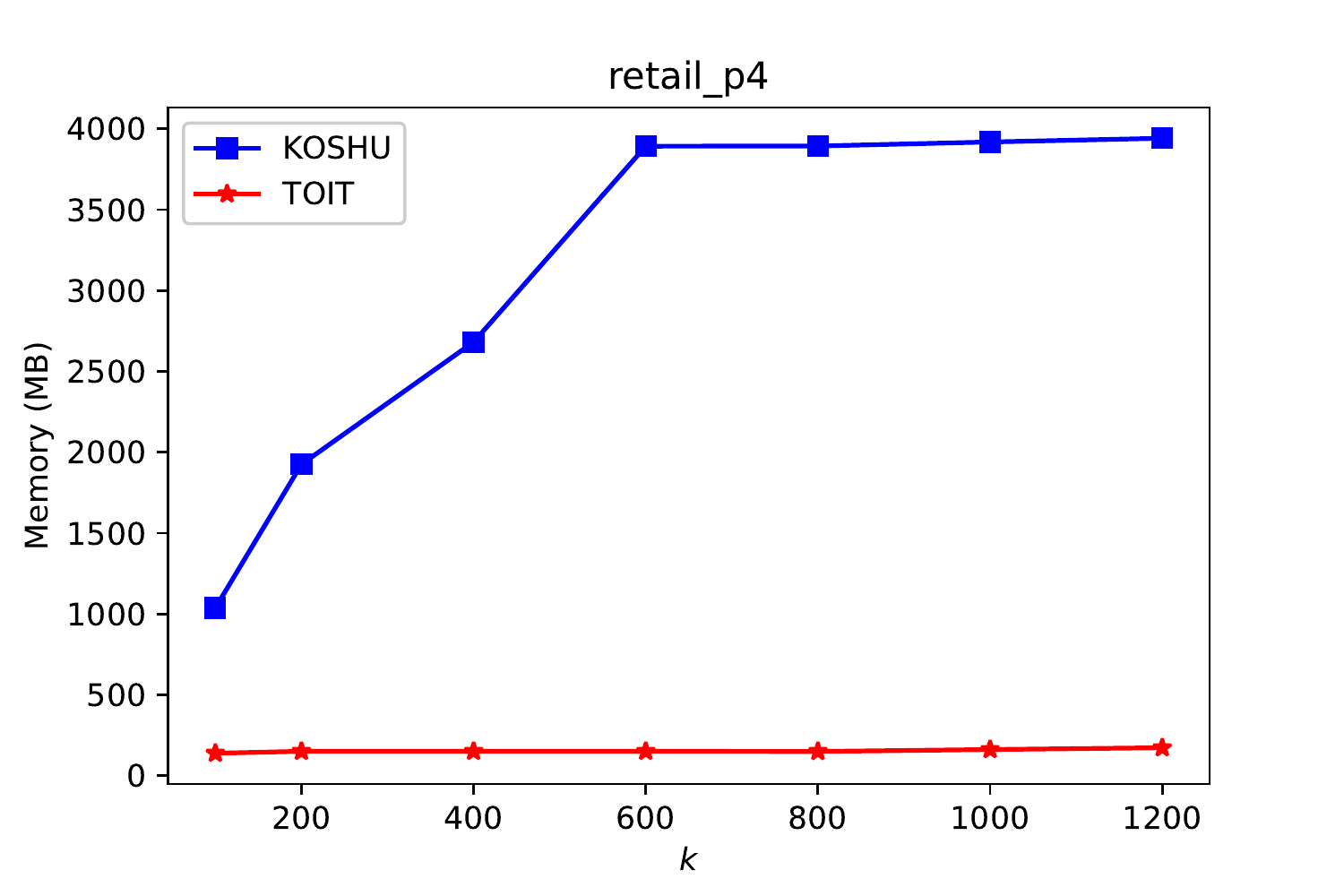}
	\end{minipage}
	\begin{minipage}[t]{0.3\textwidth}   	
		\centering   	
		\includegraphics[scale=0.35, trim=50 0 0 0]{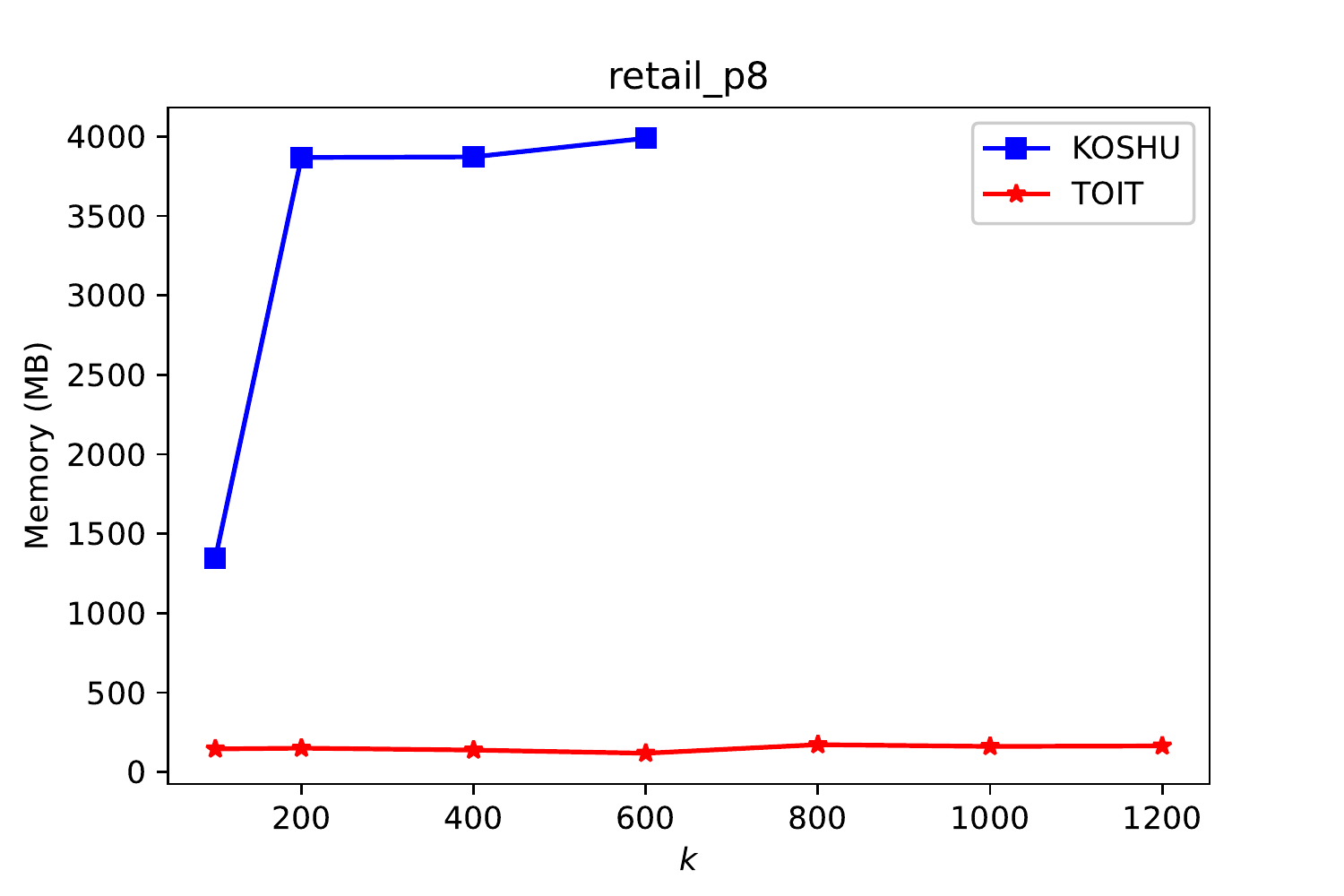}   	
	\end{minipage}
	\begin{minipage}[t]{0.3\textwidth}   	
		\centering   	
		\includegraphics[scale=0.35]{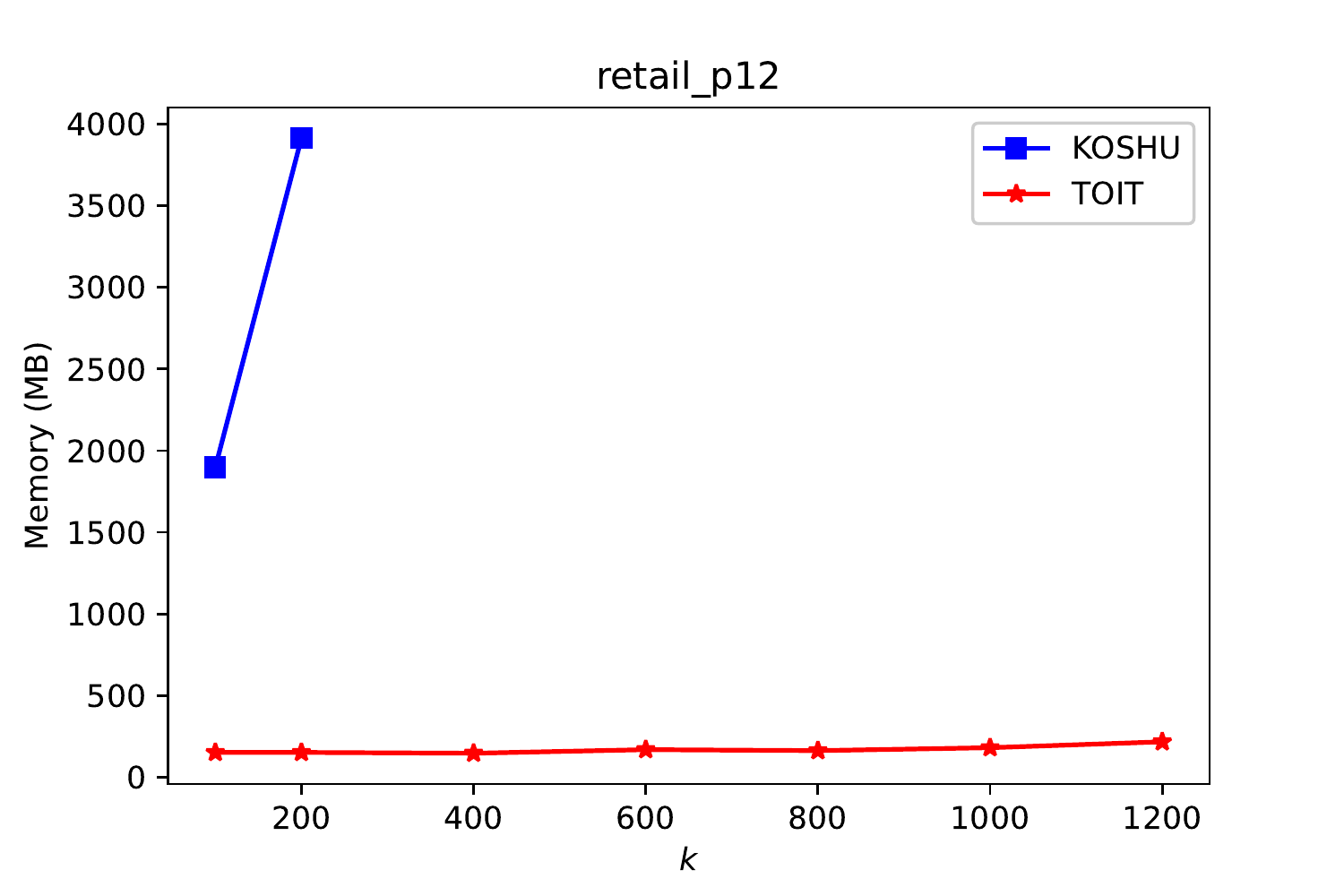}
	\end{minipage}
	
	\caption{Memory cost on different datasets with various $k$}
	\label{fig:memory}	
\end{figure}

\subsection{Influence of the Top-$k$ Threshold on Memory Cost}

In this subsection, we take the term of memory consumption experiments into account. As shown in Fig. \ref{fig:memory}, the increment in memory cost of KOSHU is noticeable, but not that of TOIT. In other words, TOIT performs much better than KOSHU with various $k$ and time periods. In all tested datasets, the memory consumption of TOIT is about 20x to 30x smaller than that of KOSHU. Because of the inefficient data structure, KOSHU takes more than 10,000 MB of memory even if $k$ is nearly 0. On chess, when $k$ is 3,000 and time period is 12, KOSHU costs nearly 10,000 MB of memory. On mushroom, in any case, the upper bound of memory consumption for TOIT is never more than 500 MB. It is a good performance that demonstrates the efficiency of our proposed algorithm. At the same time, TOIT consumes around 200 MB memory. Moreover, on the retail dataset, the memory consumption of KOSHU increases sharply while $k$ is over 100.

We suppose the reason is that KOSHU belongs to a list-base algorithm, which has to store many tuples to extend itemsets. Thus, a lot of memory is wasted. On the other hand, KOSHU utilizes poor pruning strategies, which generates a huge number of candidates, which constructs many unnecessary utility lists. This explains why the performance of KOSHU is unsatisfying in part. Instead, our novel algorithm adopts efficient database projection transaction merging technologies, which reduce the search space and the number of candidates effectively. In general, our novel algorithm performs much better than KOSHU in terms of memory consumption, especially in dense datasets.

\subsection{Influence of Different Time Periods on Execution Time and Memory}

\begin{figure}   	
	% accidents
	\begin{minipage}[t]{0.3\textwidth}	
		\centering   	
		\includegraphics[scale=0.35, trim=100 0 0 0]{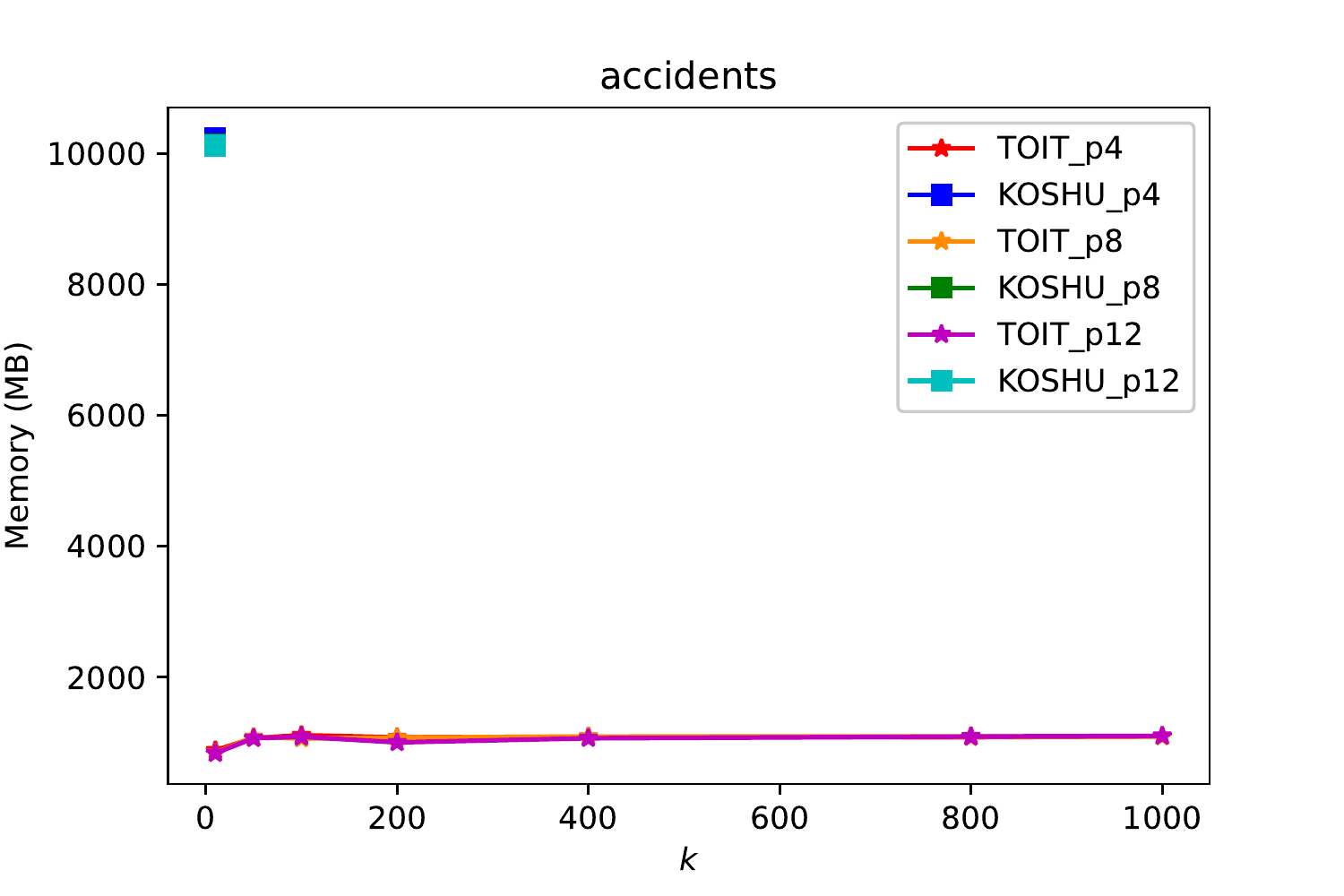}
	\end{minipage}
	\begin{minipage}[t]{0.3\textwidth}   	
		\centering   	
		\includegraphics[scale=0.35, trim=50 0 0 0]{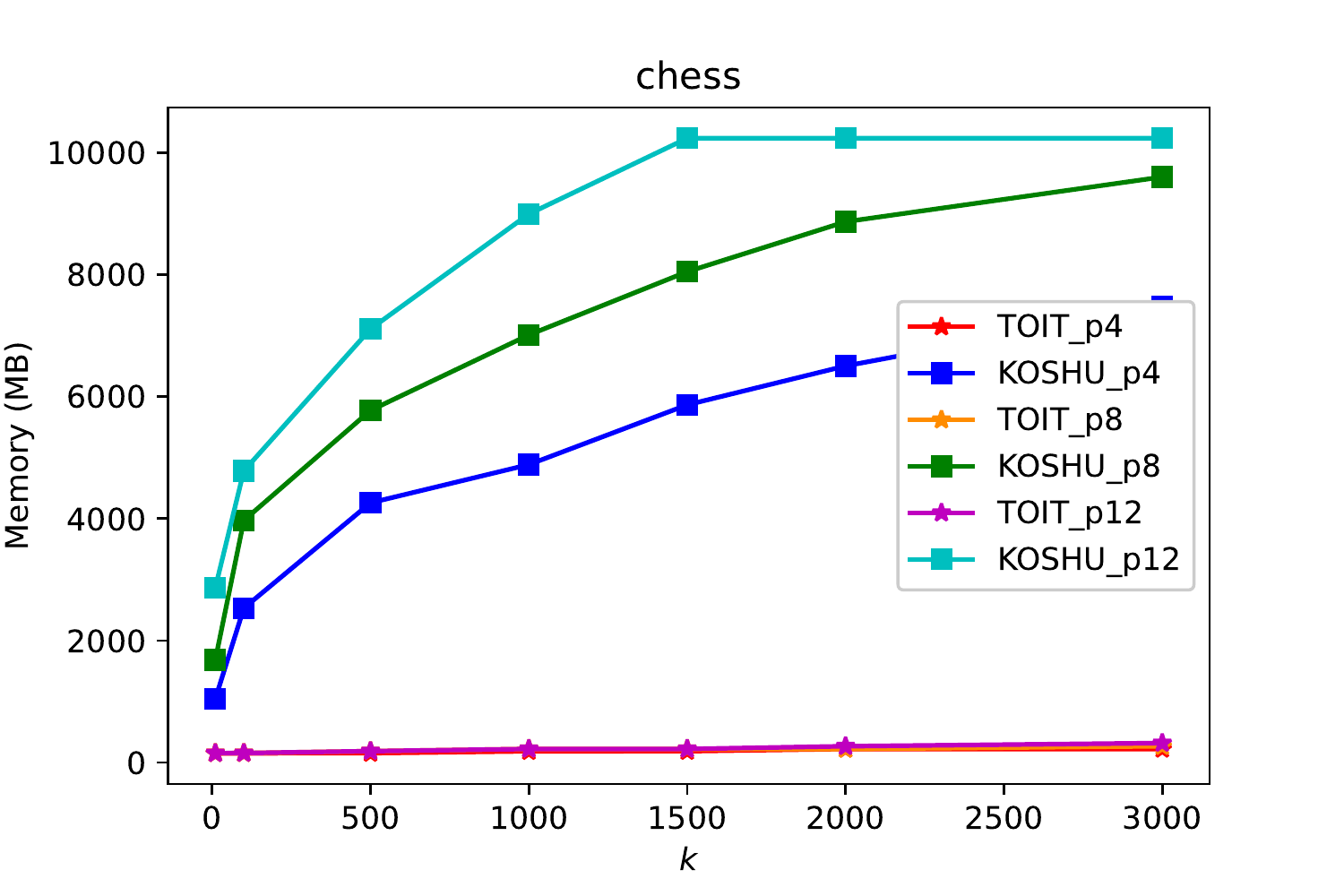}
	\end{minipage}
	\begin{minipage}[t]{0.3\textwidth}   	
		\centering   	
		\includegraphics[scale=0.35]{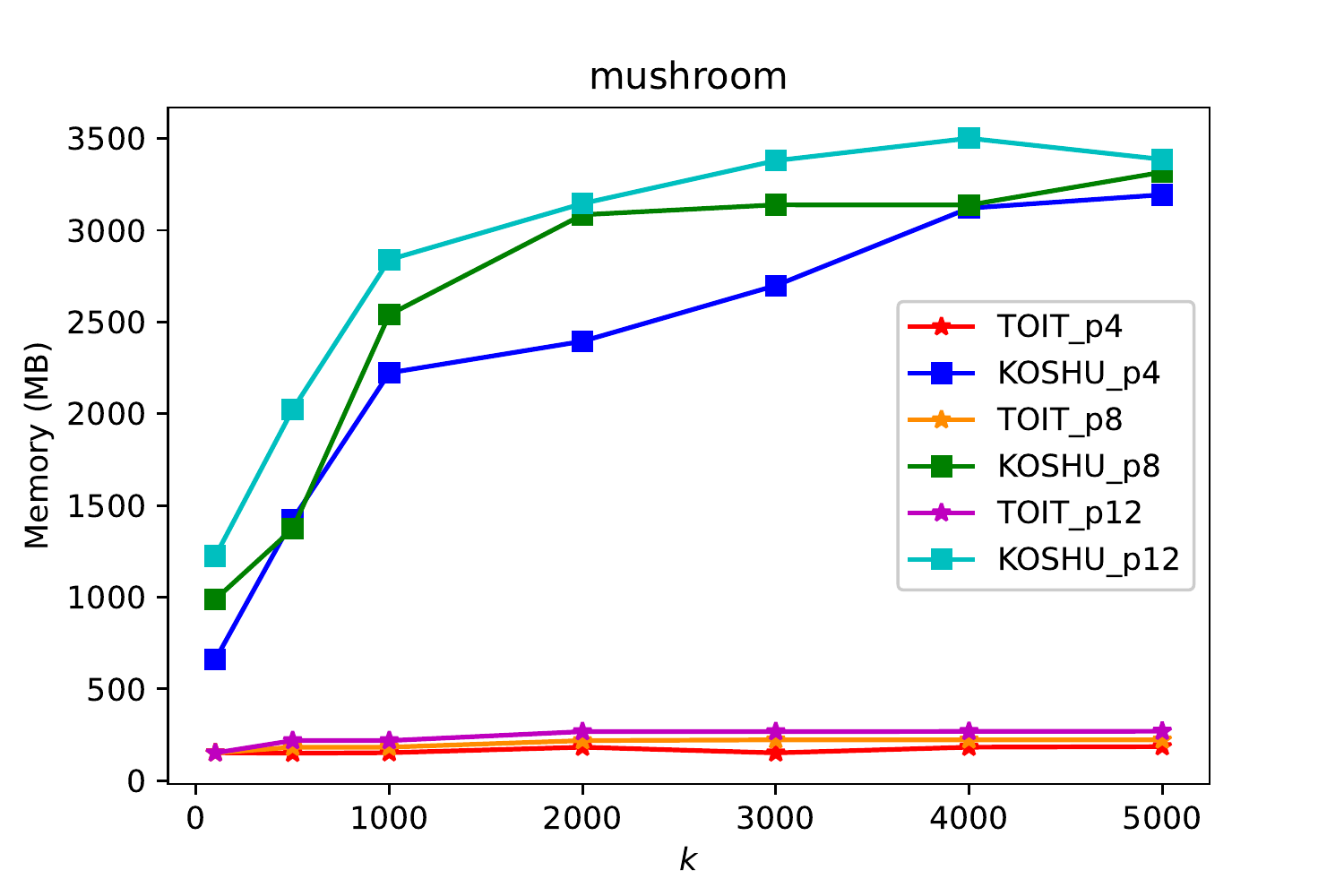}
	\end{minipage}
	
	% chess
	\begin{minipage}[t]{0.3\textwidth}	
		\centering   	
		\includegraphics[scale=0.35, trim=100 0 0 0]{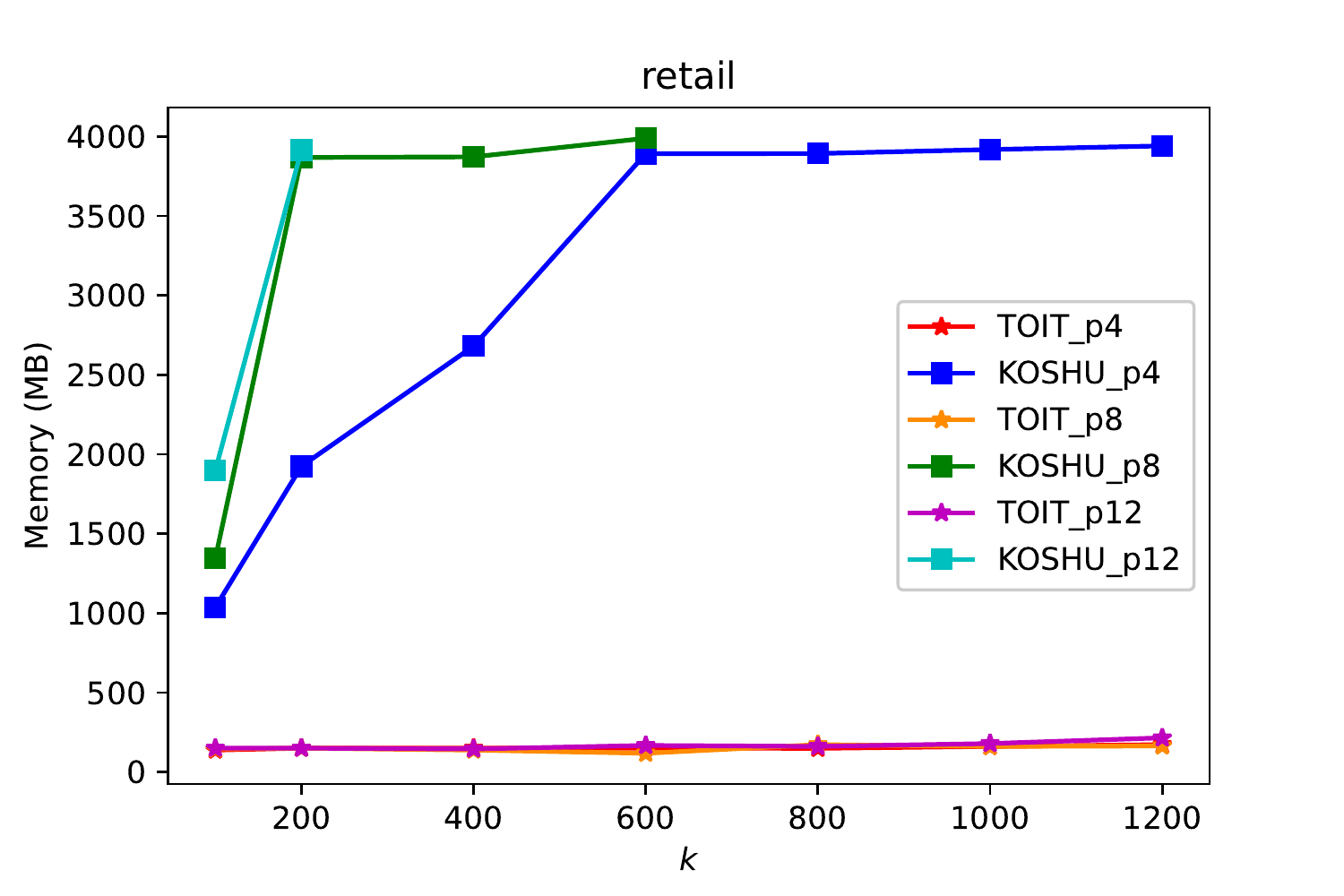}
	\end{minipage}
	\begin{minipage}[t]{0.3\textwidth}   	
		\centering   	
		\includegraphics[scale=0.35, trim=50 0 0 0]{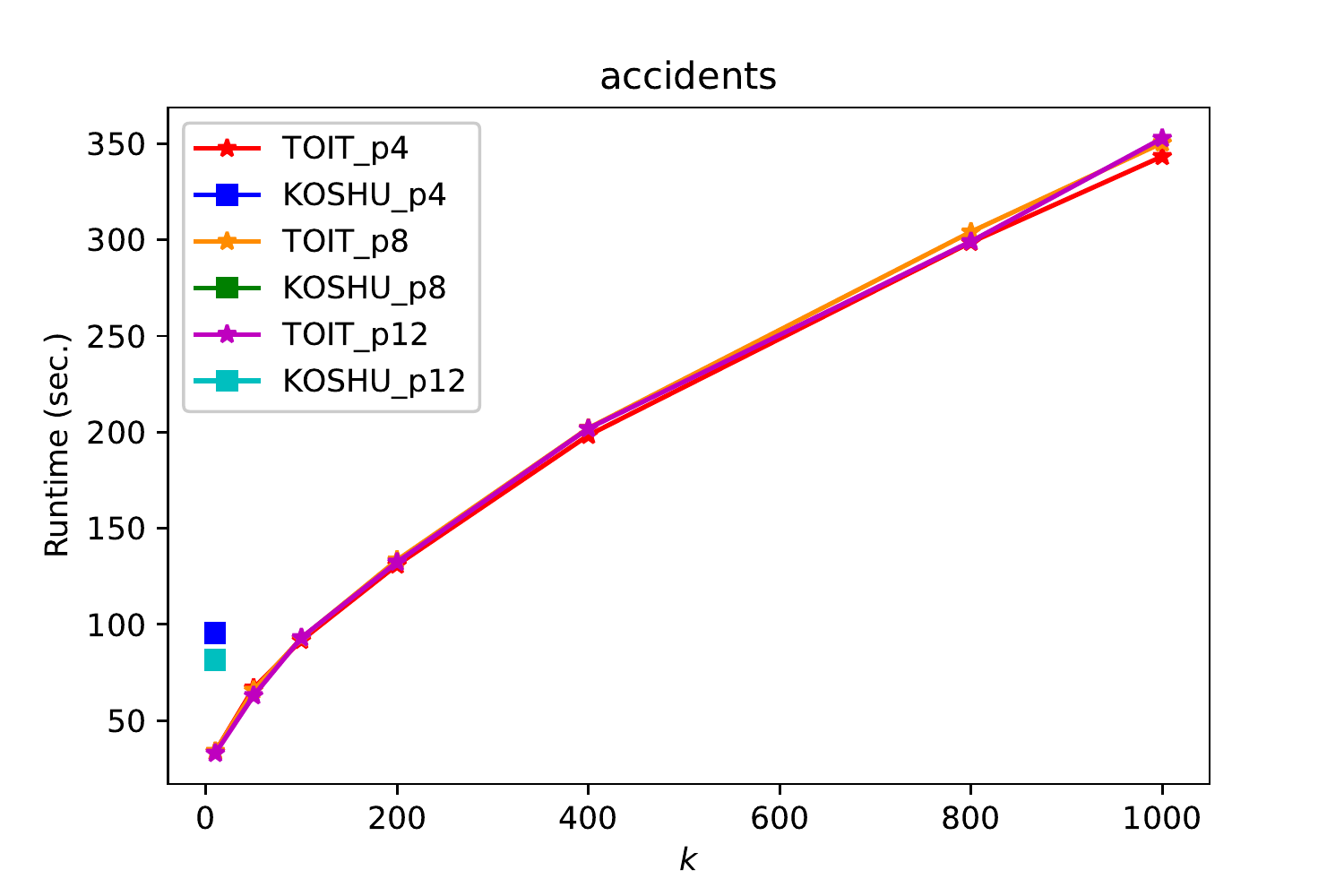}   	
	\end{minipage}
	\begin{minipage}[t]{0.3\textwidth}   	
		\centering   	
		\includegraphics[scale=0.35]{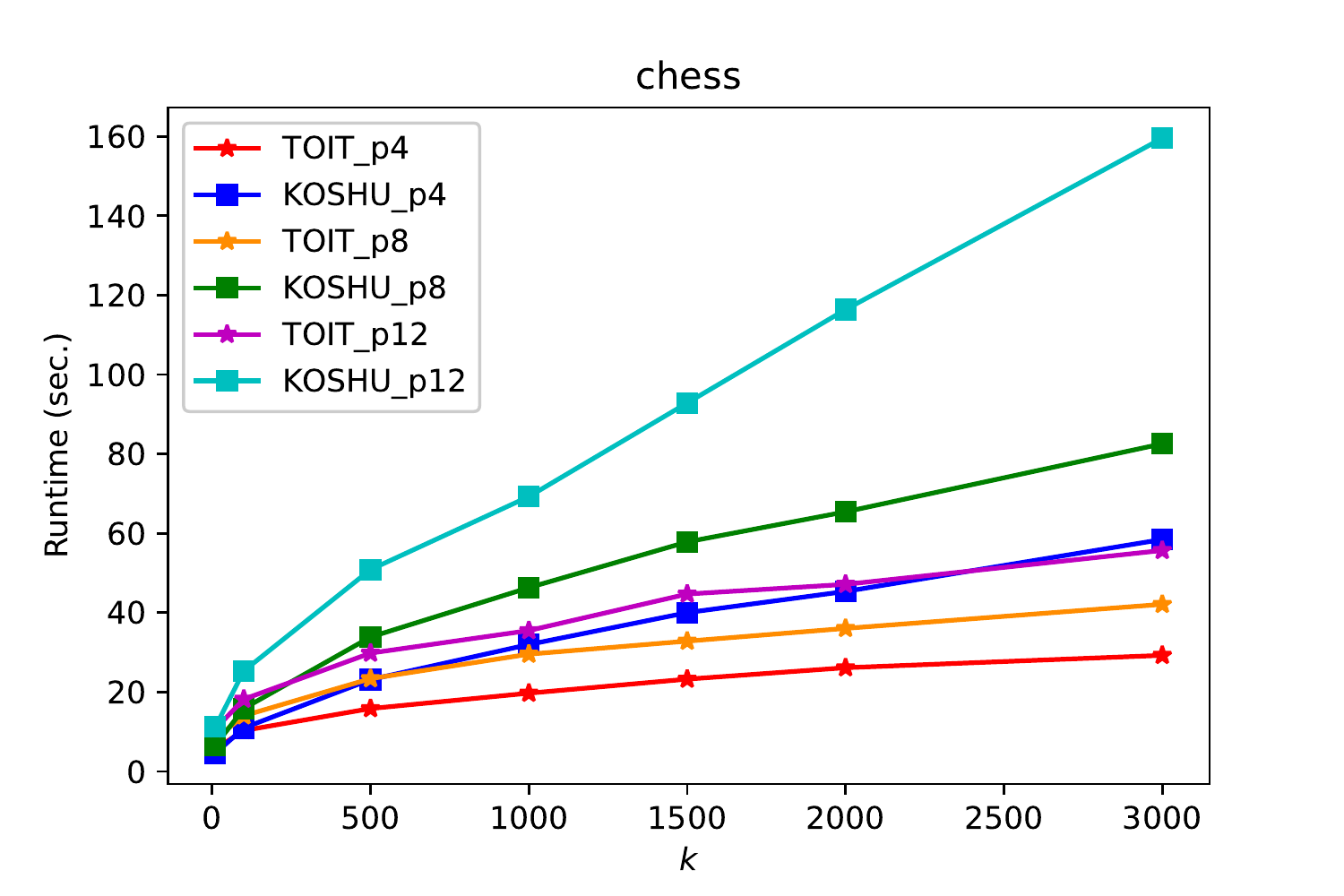}
	\end{minipage}
	
	\begin{minipage}[t]{0.3\textwidth}	
		\centering   	
		\includegraphics[scale=0.35, trim=100 0 0 0]{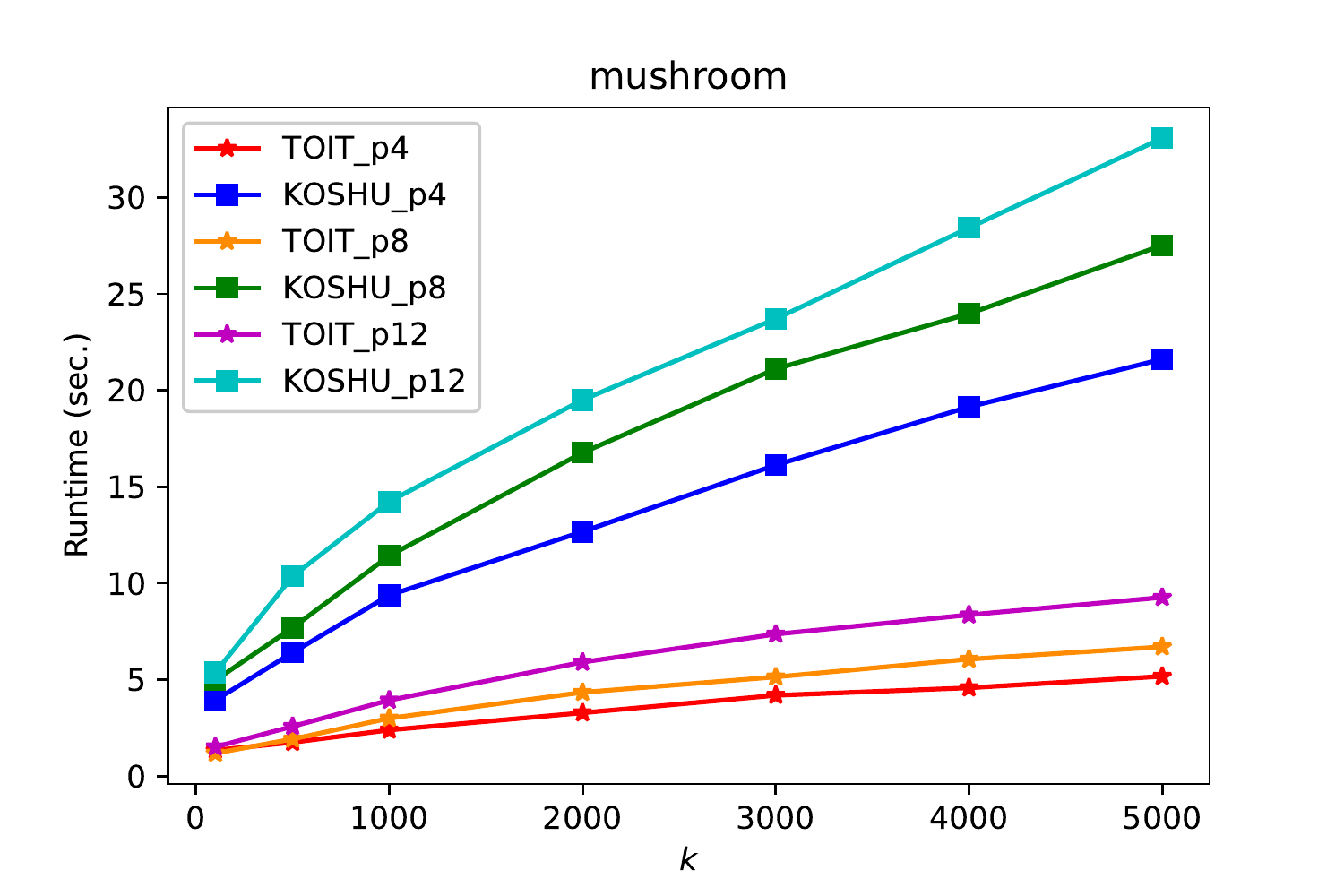}
	\end{minipage}
	\begin{minipage}[t]{0.3\textwidth}	
		\centering   	
		\includegraphics[scale=0.35]{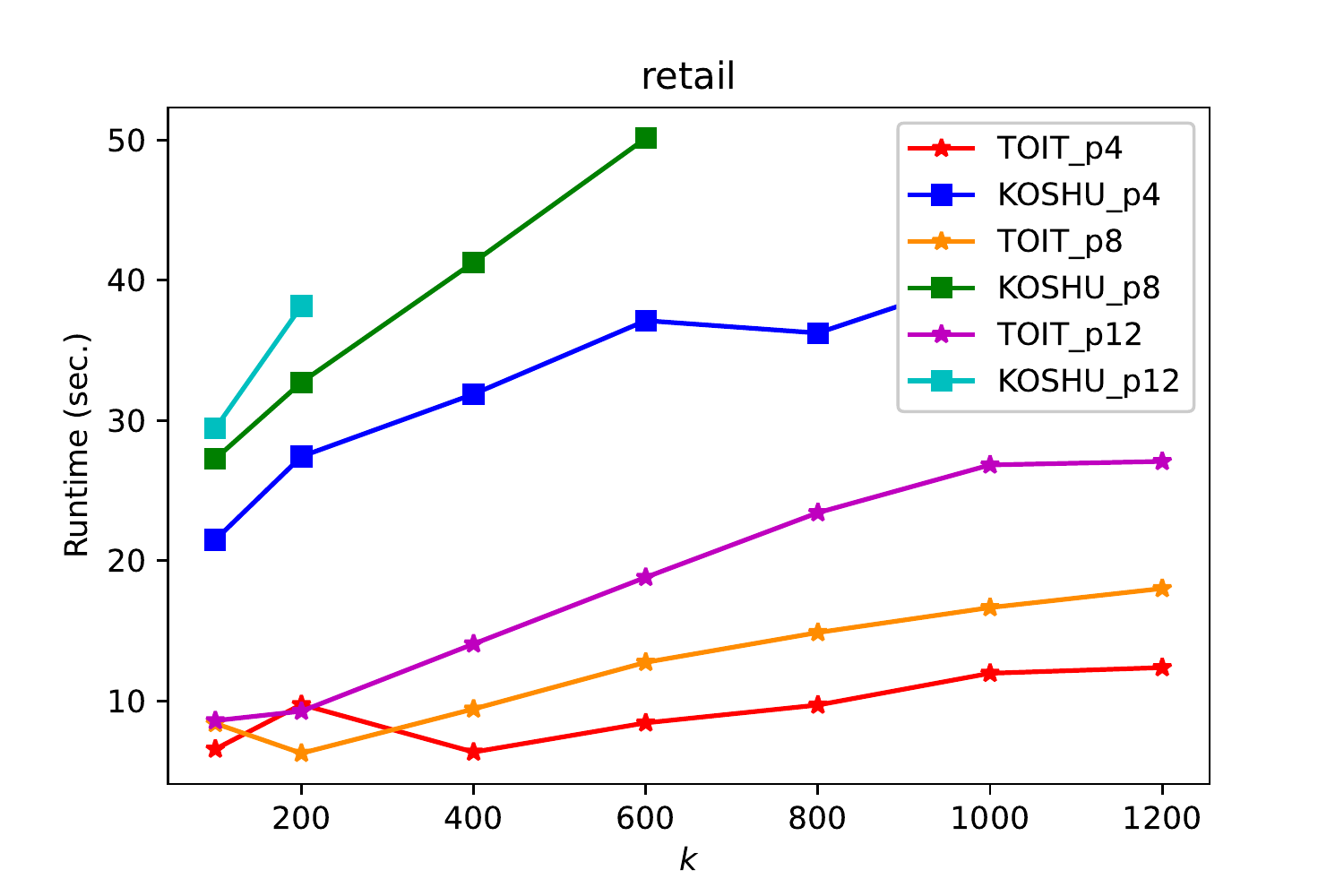}
	\end{minipage}
	
	\caption{Memory and time cost on different periods under changed $k$}
	\label{fig:memorycompared}	
\end{figure}

As shown in Fig. \ref{fig:memorycompared}, three time periods (i.e., 4, 8, and 12) are considered in accidents, chess, mushroom, and retail datasets, respectively. We will analyze the experimental results to discuss the influence of different time periods on runtime and memory consumption. In accidents, the execution time cost of TOIT raises slightly when the time period increases from 4 to 12. However, TOIT uses almost the same memory consumption in four datasets. This shows our new algorithm is able to save memory effectively and decrease the number of useless candidates during the mining process. According to KOSHU, it is clear that as time passes, not only runtime costs but also memory consumption increase rapidly. It can be predicted that when the time period is longer than 12, the runtime and memory consumption of KOSHU will be unacceptable. To summarize, the memory cost of TOIT appears more stable than that of KOSHU, and the runtime cost of TOIT is also acceptable.

\begin{table*}[!ht]
	\caption{Discovered patterns under changed $k$}
	\begin{minipage}[t]{\textwidth}	
		\centering   	
		\includegraphics[scale=0.8]{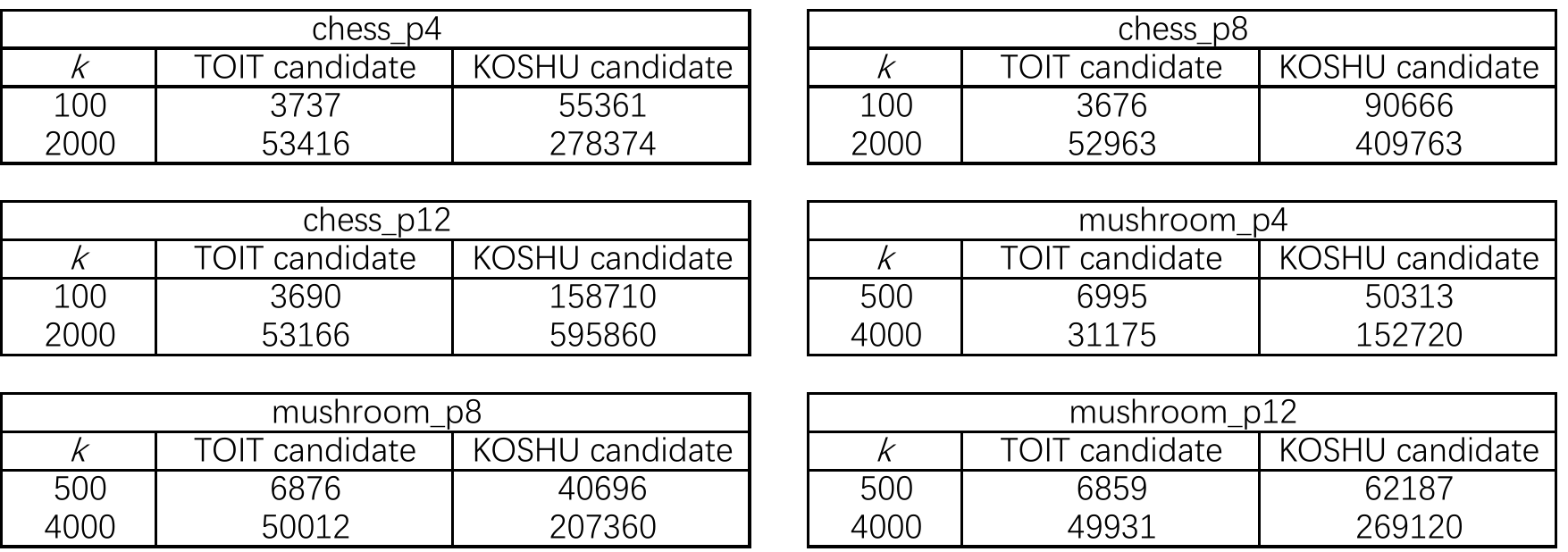}
		\label{fig:candidates}
	\end{minipage}
\end{table*}

\subsection{Influence of Different Time Periods on Discovered Patterns}

We also study the relationship between different time periods and the number of discovered candidates. In Table \ref{fig:candidates}, we tested two algorithms with three different time periods (i.e., 4, 8, and 12) in chess and mushroom datasets. Obviously, the number of candidates for KOSHU can be one order of magnitude greater than that of TOIT. For example, when $k$ is 4,000 and the time period is 8, KOSHU generates 207,360 candidates on mushroom, which is 4x more than that of TOIT. Considering the chess dataset, when the time period is 4, 8, and 12, the number of candidates generated by TOIT is nearly unchanged. This is because $k$ is very small and chess has the fewest distinct items in four datasets. As a result, effective pruning strategies will reduce a large number of unpromising items during mining. In a word, the TOIT algorithm has good performance in candidate generation.

\section{Conclusions and Future Work} %  and Future Work
\label{sec:conclusion}

In this paper, we propose a generic algorithm to solve the task of mining top-$k$ on-shelf high utility patterns efficiently. TOIT is a one-phase algorithm that searches the on-shelf database in depth for patterns with both positive and negative values. To improve mining efficiency, TOIT applies the RIU strategy to raise the internal utility by comparing it to the utility of a single item. It utilizes database projection and transaction merging to avoid scanning the large original dataset frequently. Furthermore, two effective upper bounds, named local utility and subtree utility, are used to reduce the search space. We perform comparative experiments with the state-of-the-art KOSHU algorithm on several datasets with various characteristics. The experiment results show that TOIT outperforms KOSHU both in memory consumption and executing time. In the future, we aim to improve the TOIT algorithm on both sides. On the one hand, we will explore more efficient strategies, which in terms of increasing internal utility as early as possible; on the other hand, we will try to extend this mode to deal with the pattern discovery task in more types of datasets in other areas, such as dynamic data mining, fuzzy mining, and sequence mining.

\begin{acks}

This research was supported in part by the National Natural Science Foundation of China (Grant Nos. 61902079 and 62002136), Natural Science Foundation of Guangdong Province of China (Grant No. 2022A1515011861), Guangzhou Basic and Applied Basic Research Foundation (Grant Nos. 202102020928 and 202102020277).
\end{acks}

\bibliographystyle{ACM-Reference-Format}
\bibliography{paper}

\end{document}